\documentclass[twocolumn]{aastex631}
\hypersetup{linkcolor=red,citecolor=blue,filecolor=cyan,urlcolor=blue}
\usepackage[T1]{fontenc} 
\usepackage{ae,aecompl}
\usepackage{graphicx}	
\usepackage{amsmath}
\usepackage{xspace}
\usepackage{color}
\usepackage{txfonts}
\usepackage{hyperref}
\usepackage{orcidlink}
\newcommand{\pcm}{\,$\mathrm{cm^{-2}}$}	
\newcommand{\gcmq}{\,$\mathrm{G~cm^{3}}$} 
\newcommand{\cmsq}{\,$\mathrm{cm^{2}}$ }
\newcommand{\musec}{\,$\mathrm{\mu\,s}$ }

\newcommand{\kev}{\,$\mathrm{keV}$ }
\newcommand{\kpc}{\,$\mathrm{kpc}$ }
\newcommand{\ksec}{\,$\mathrm{ksec}$ }

\newcommand{\ie}{\textit{i}.\textit{e}. }

\newcommand{\xspec}{\textsl{XSPEC}\xspace}	

\def \igrj {\mbox{IGR~J16358$-$4726}}
\def \gx {\mbox{GX~301$-$2}}
\def \xper {\mbox{X~Per}}
\def \twoS {\mbox{2S~0114+650}}
\def \fourU {\mbox{4U~2206+54} }
\def  \xtej {\mbox{XTE~J1739$-$302} }

\def\swift {\emph{Swift} }

\def\xte {\emph{RXTE} }

\def\integral {\emph{INTEGRAL} }
\def\bepposax {\emph{BeppoSAX} }
\def\suzaku {\emph{Suzaku} }
\def\uhuru {\emph{Uhuru} }
\def\nustar {\emph{NuStar} }
\def\astrosat {\emph{AstroSat} }
\def\hxmt {\emph{Insight--HXMT} }
\def\sxt{\emph{SXT} }
\def\laxpc{\emph{LAXPC} }
\def\czti{\emph{CZTI} }
\def\flux {\mbox{erg cm$^{-2}$ s$^{-1}$}}
\def\lum {\mbox{erg s$^{-1}$}}

\def\arcmin {\hbox{$^\prime$}}
\def\arcsec {\hbox{$^{\prime\prime}$}}
\def\chiq {$\chi_{\nu}^{2}$}
\def\chisq {$\chi^{2}$}

\def\ktbb{kT_{\rm bb}}
\def\kt0{kT_{\rm 0}}
\def\kte{kT_{\rm e}}
\def\r0{r_{\rm 0}}

\def \tbabs {\textsc{tbabs}}
\def \tbpcf {\textsc{tbpcf}}

\def \cpl {\textsc{cutoffpl}}
\def \pl {\textsc{powerlaw}}
\def \hcp {\textsc{highecut}}
 
\def \comptt {\textsc{comptt} }
\def \compmag {\textsc{compmag} }

\def\b0{\beta_{\rm 0}}

\begin{document}

\title{\astrosat and \hxmt observations of the long-period X-ray pulsar \fourU}
\label{firstpage}
\author{Prahlad R. Epili\,\orcidlink{0000-0002-2278-8375}}
\affiliation{Department of Astronomy, School of Physics and Technology, Wuhan University, Wuhan 430072, China P.R.C}
\altaffiliation{Email address: prahlad@whu.edu.cn, wangwei2017@whu.edu.cn}
\author{Wei Wang\,\orcidlink{0000-0003-3901-8403}}
\affiliation{Department of Astronomy, School of Physics and Technology, Wuhan University, Wuhan 430072, China P.R.C}

\begin{abstract}	

We report the timing and spectral studies of the accreting X-ray pulsar \fourU using \astrosat and \hxmt observations 
taken in 2016 and 2020 respectively. X-ray pulsations from the system are detected by both missions. The \astrosat discovered a significant periodic signal at $\sim 5619$ s in 2016 and \hxmt found a pulsation period at $\sim 5291$ s 
in 2020.  A comparison of its spin-period evolution with the present spin-period estimates shows that the neutron star in \fourU now has recently undergoing a spin-up episode after attaining to its slow pulsations of ~5750~s around 2015 from its prolonged spin-down phase. The present average spin-up rate of the pulsar is found to be at $\sim1.2\times10^{-13}$ Hz~s$^{-1}$. The phase-averaged spectra of the pulsar in 1-60\kev could be explained with a high energy cutoff power-law continuum model, no evident line features are found with \astrosat. The application of 
Comptonization models such as \comptt and \compmag to the phase averaged spectra of \fourU reveal a hotter source photon region near the pulsar with an emission size extending to $\sim 2-2.8$~km. Using the quasi-spherical settling accretion theory, we explain the present spin-up and the possibility of strong magnetic field of the pulsar.

\end{abstract}


\section{Introduction} \label{sec:intro}
\par
\fourU is a persistent low luminosity ($\sim10^{35}$\lum) high mass X-ray binary (HMXB) system consisting of a neutron star 
orbiting an O-type companion star. The neutron star in this system is known to show long period pulsations of
$\sim5550\sec$ \citep{2009A&A...494.1073R,2010ApJ...709.1249F,2010A&A...520A..22W,2013MNRAS.432..954W}. Along with \twoS, \gx, 
\igrj\ and \xper\  it belongs to a peculiar family of HMXB sources \citep{2012int..workE..12B,2014mbhe.conf..215B} in which the 
NS pulse period is longer than 100s of seconds, has a persistent low luminosity, accretes from the 
slow and dense stellar wind of companion star and NS shows long-term spin down trend since 2000 \citep{2013MNRAS.432..954W}.

\fourU remains till date, an exceptional member of high mass X-ray binary systems. Although it was identified as an X-ray source 
since more than 20 years ago, the nature of the compact object remained uncertain until \citet{2009A&A...494.1073R} found 
a $\sim5560~s$ modulation in the X-ray light curve of the source. It was later understood as spin period of the neutron star.
The findings of slow pulsations have put an end to the long standing debate on the nature of compact object in \fourU. 
However, it posed new challenges in classifying the source in either subgroups of HMXBs. The optical companion in \fourU is a 
main-sequence O9.5V star having excess in He abundance which suggests against the supergiant X-ray binary classification.
The X-ray light curve of \fourU also differs from majority of Be/X-ray binaries \citep{2012MNRAS.425..595R} in a way that 
there are no Type-I X-ray outbursts or any aperiodic giant Type II outbursts (without any modulations on the orbital period). 
The X-ray emission from the source is understood to be due to accretion of slow and dense wind from the non-supergiant companion star. 
As per spin-evolutionary models in close binaries, the slow pulsations found in 4U~2206+54 imply its surface magnetic field value of 
the order $\ge10^{14}$ G. This would lead to assumption that the X-ray source in \fourU binary system is an accreting magnetar 
\citep{2010ApJ...709.1249F}. 

Since its discovery with \uhuru in 1972, \fourU is observed with various X-ray observatories. From the earliest $\sim9$~ks 
\xte~PCA observations of \fourU (i.e. March 1997), it was reported that the temporal X-ray variability of the source is 
associated with flaring activity on short time scales \citep{2001A&A...371.1056N}. The X-ray intensity was varying by a 
factor of 3 within 2 minutes. \fourU was also caught in a hard X-ray outburst during December~2005 with {\it INTEGRAL/IBIS}. 
The hard X-ray outburst had double-flare events lasting for 2-days. The first flare was seen with a peak luminosity of 
$\sim4\times10^{36}$~erg~s$^{-1}$ in 1.5-12 keV range whereas the second flare had an average X-ray luminosity of 
$1.3\times10^{36}$ erg~s$^{-1}$ in 20-150 keV energy band \citep{2010A&A...520A..22W}. The reported bright flares are 
similar to those found in SFXTs but the physical mechanisms responsible for sudden and enhanced wind accretion in \fourU 
remains speculative. It was further suggested  that the accreting material in the form of clumps from the stellar wind 
could hit the polar cap region of neutron star in \fourU leading to hard X-ray outbursts. Hence a detailed study is 
required to understand the various effects on accretion through the capture of stellar wind of a main sequence star in 
\fourU and thereby providing the missing link between these systems and  SFXTs \citep{2010A&A...520A..22W}.  

The evolutionary trend of slow pulsation period ({\it i.e.} longer than 5000~s) and origin of X-ray variability in \fourU 
remain poorly understood. Since \fourU is one of the weak and low-luminosity hard X-ray source and the available observations 
are relatively short, there exists a large uncertainty in determining the spin period and period derivative for the pulsar. 
Long pulsation period ($\sim 5560~s$) of \fourU makes it one of the accreting X-ray pulsar with slowest pulsations.
According to standard models of spin-period evolution of young neutron star in close binary systems, the new-born neutron 
star goes through three primary stages of evolution \citep{1991PhR...203....1B} before being switched on as an X-ray pulsar.
In the first stage  {\ie \it ejector state}, the neutron star loses rotational energy through spinning down. 
In the {\it propeller state}, the spin period of the neutron star decreases due to interaction between its magnetosphere 
and the stellar wind of the companion. In {\it accretor state}, the neutron star starts accreting material on to the surface 
when its spin period reaches a critical value. From then onwards, it beams X-ray emissions as an X-ray pulsar. The spin period of the 
neutron star during {\it ejector} and {\it propeller} states could only increase to the longest period of several hundred
seconds but cannot be more than 1000~s \citep{1998MNRAS.299...73U, 2013MNRAS.432..954W}. 
Hence, the very slow pulsations of period longer than 5000~s found in \fourU could not be explained in this standard picture 
\citep{2013MNRAS.432..954W}. So, the evolution pathway of these slow pulsar is still unknown and remains quiet an 
interesting topic in astrophysics.

Detection of cyclotron lines in the spectra of accretion powered X-ray pulsars gives us a direct estimate of neutron star 
surface magnetic field. In \fourU, a tentative detection of cyclotron resonance scattering feature (CSRF) near 
$\sim30$\kev has been reported in past by \xte, \bepposax and \integral observations at different epochs between 1997
and 2005 \citep{2004A&A...423..301T,2009MNRAS.398.1428W}. However there was no sign of any CRSF detected from a long 
\xte observation ($\sim$137~ks exposure) in May~2007 \citep{2009A&A...494.1073R}. From the \bepposax and \suzaku observations, 
the spin-down rate of the pulsar was reported to be ${\dot\nu}=(-1.5\pm0.2)\times10^{-14}$~Hz~$s^{-1}$ \citep{2010ApJ...709.1249F}.
which would imply a magnetic field of $B\sim10^{14}~G$ and hence the corresponding electron cyclotron line would appear at around 
$E > 500~keV$ which is beyond the range of available detectors. However, the estimated spin-down rate is based only on two 
observations of the pulsar. In our present work, we have discussed the \astrosat and \hxmt observations of \fourU taken 
in September 2016 and October 2020 respectively. In brief, the paper is structured as follows: in Section~\ref{sec:obs}, 
we present the observation details. In Section~\ref{sec:timing}, we have analysed the light-curves of \fourU in various 
energy ranges. Subsequent spin-period estimation and pulse-profiles studies have been carried out. The spectral study, 
in particular the phase-averaged spectroscopy from \astrosat observations of \fourU has been presented in 
Section~\ref{sec:spectra} with the best-fitted empirical and physics-based models. And lastly in Sections~\ref{sec:discuss} 
and \ref{sec:summary}, we discuss the spectral parameter variations in the prevailing theoretical frameworks and present the 
conclusions, respectively.

\section{Observations and Data Analysis} \label{sec:obs}
\subsection{\astrosat}
\astrosat \citep{2014SPIE.9144E..1SS} is India's first space observatory capable of observing the stellar sources in 
multi-wavelengths. It has been launched on September 28, 2015 carrying onboard five sets of scientific instruments. 
These are namely, a Soft X-ray Telescope (\sxt, \citealt{2016SPIE.9905E..1ES,2017JApA...38...29S}), the Large Area 
X-ray Proportional Counters (\laxpc, \citealt{2017JApA...38...30A,2016SPIE.9905E..1DY}), a Cadmium Zinc Telluride Imager 
(\czti, \citealt{2017JApA...38...31B,2017CSci..113..595R}), a Scanning Sky Monitor (SSM, \citealt{2017JApA...38...32R,
2017ExA....44...11R}) and an Ultraviolet Imaging Telescope (UVIT, \citealt{2017AJ....154..128T}). Also there is a Charged 
Particle Monitor (CPM, \citealt{2017JApA...38...33R}) onboard \astrosat to monitor the charged particle count rate variation. 
As the satellite passes through the SAA region, the CPM helps in optimizing the operation time of the X-ray detectors and effectively 
switching off these instruments during SAA passage to avoid the adverse effects on the X-ray detectors and their subsequent 
saturation. 

The array of LAXPC detectors onboard \astrosat covers an energy range of 3-80~keV with a large collection area of 
$\sim6000$~\cmsq at 15~keV \citep{2016SPIE.9905E..1DY}. 
It consists  of 3 independent and identical units of proportional counter detectors 
(LAXPC-10, 20 \& 30) working simultaneously in different modes. The default operation modes of LAXPC are: 
(a) Broad Band Counting of data (i.e BB mode) and (b) Event Analysis mode data (i.e EA mode). 
The Soft X-ray Telescope (SXT, \citealt{2016SPIE.9905E..1ES}) onboard \astrosat is sensitive to soft X-rays in 
0.3 -- 8 keV energy range.
The total field of view (FOV) of SXT is $\sim40\arcmin$ with 
an effective area that varies from $\sim128$\cmsq at 1.5 \kev to $\sim20$\cmsq at 6~\kev. The hard X-ray coverage 
onboard \astrosat is provided with the Cadmium Zinc Telluride Imager (\czti, \citealt{2017CSci..113..595R, 
2017JApA...38...31B}) in 20-200\kev. Its a 2D coded mask imager with The coded aperture mask on \czti\ with a 
field of view of $4.6\deg \times 4.6\deg$ (FWHM) provides 8\arcsec resolution for bright sources. The energy resolution 
of \czti is of $\sim11\%$ at 60 \kev.

\begin{figure*}
 \begin{center}
 \includegraphics[scale=0.3]{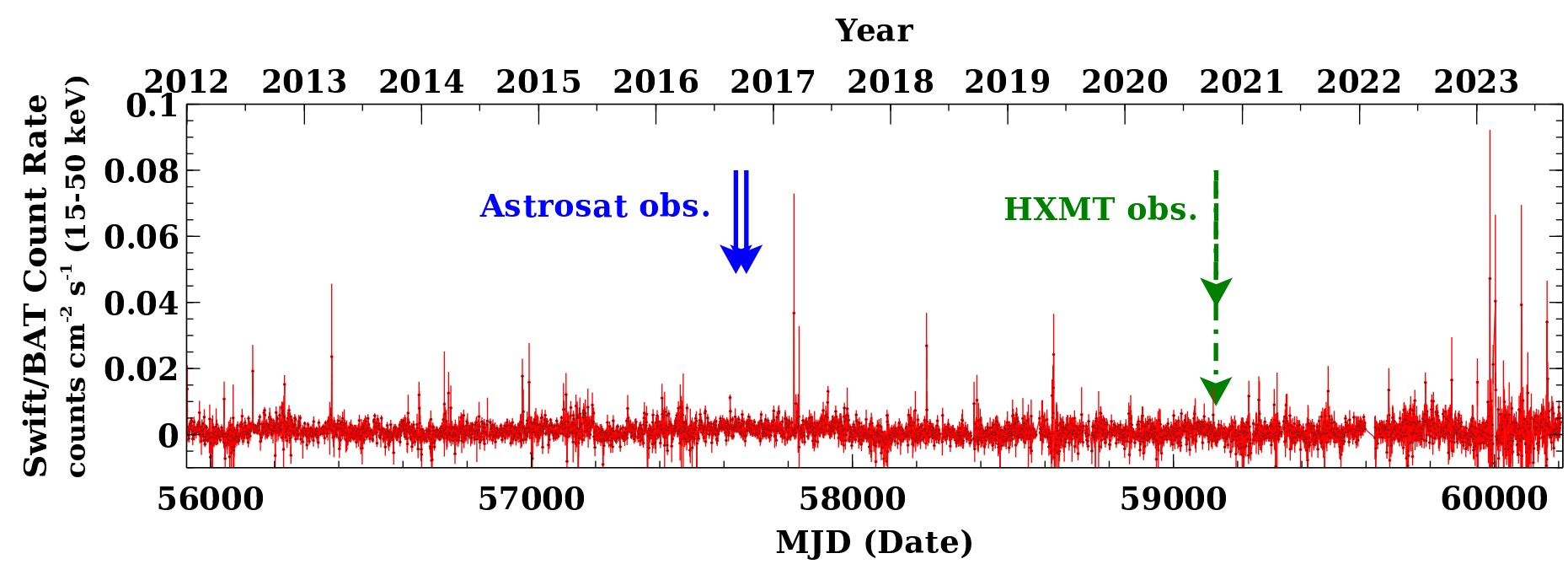}
 \end{center}
 \caption{Swift/BAT light curve of \fourU in 15--50\kev. The arrows indicate the respective epochs  of \astrosat and \hxmt observations used in the present work.} 
 \label{fig:1}
 \end{figure*}

\fourU has been observed with \laxpc, \sxt, \& \czti instruments onboard \astrosat on two occasions (\ie on 06-09-2016
and 08-10-2016) respectively with observation IDs: 9000000644, 9000000720. The effective exposure of the first 
observation with \laxpc\ is $\sim$43.7\ksec. Whereas the  second observation has an effective exposure of $40.7$\ksec 
with \laxpc detectors. The observations are also carried simultaneously with \sxt \& \czti\ instruments. A log of these 
observations used in our study are given in Table.~\ref{table:1}. In Figure~\ref{fig:1}, we show the long-term \swift/BAT 
light curve of \fourU in 15-50 \kev. The epochs of \astrosat observations of \fourU used in the paper are shown with 
arrows in the figure.  In the present work we have used the data from both of these observations to study the X-ray 
spectrum in  $1-60$~\kev.  For the X-ray timing studies we have used the \astrosat observations in energy range: 0.3--5\kev 
from \sxt, 5--30\kev from \laxpc and 30--60\kev from \czti observations of \fourU respectively. 
\begin{table*}
\begin{center}
\centering 
\caption{Summary of the X-ray observations of \fourU.} 
\label{table:1}
\begin{tabular}{@{}lcccccc}
\hline
Satellite &Obs.\,ID & Obs.~Date & Instrument    & Effective Exposure  &Count-rate & Observation Mode$^{a}$\\
&  & & & (ksec) &($cs^{-1}$) & \\
\hline 
\astrosat  &9000000644 & 2016 Sept 6  & SXT   & 21.4 &$\sim0.55$ & PC ($\sim2.38$\,s)  \\
           &           &              & LAXPC & 43.7 &$\sim35.05$ & EA  (10\,\musec)\\
           &           &              & CZTI  & 44.7 &$\sim1.24$ & M0  (20\,\musec) \\
\\
          &9000000720 & 2016 Oct 8   & SXT   & 26.1 &$\sim1.10$ & PC ($\sim2.38$\,s)  \\
           &           &              & LAXPC & 40.7 &$\sim45.84$ & EA  (10\,\musec)\\
           &           &              & CZTI  & 46.3 &$\sim0.22$ & M0  (20\,\musec)  \\
\\
\hxmt      &P0305070003 &2020 Oct 10  & LE    &32.5  &$\sim2.36$ &-- (1\,msec)  \\
           &            &             & ME    &53.6  &$\sim4.40$ &-- (240\,\musec)  \\
           &            &             & HE    &52.0  &$\sim10.13$ &-- (4\,\musec)  \\
\\
           &P0305070004 &2020 Oct 11  & LE    &14.6  &$\sim2.38$ &-- (1\,msec)  \\
           &            &             & ME    &21.8  &$\sim4.43$ &-- (240\,\musec)  \\
           &            &             & HE    &23.4  &$\sim9.76$ &-- (4\,\musec) \\
\hline
\end{tabular}
\begin{list}{}{}
\item[$^{a}$] EA: Event Analysis,  PC: Photon Counting, M0: regular event mode data. 
The detector resolution time is given in parentheses for corresponding observation mode.
\end{list}
\end{center}
\end{table*}

\begin{figure}[!ht]
    \includegraphics[height=5cm,angle=0]{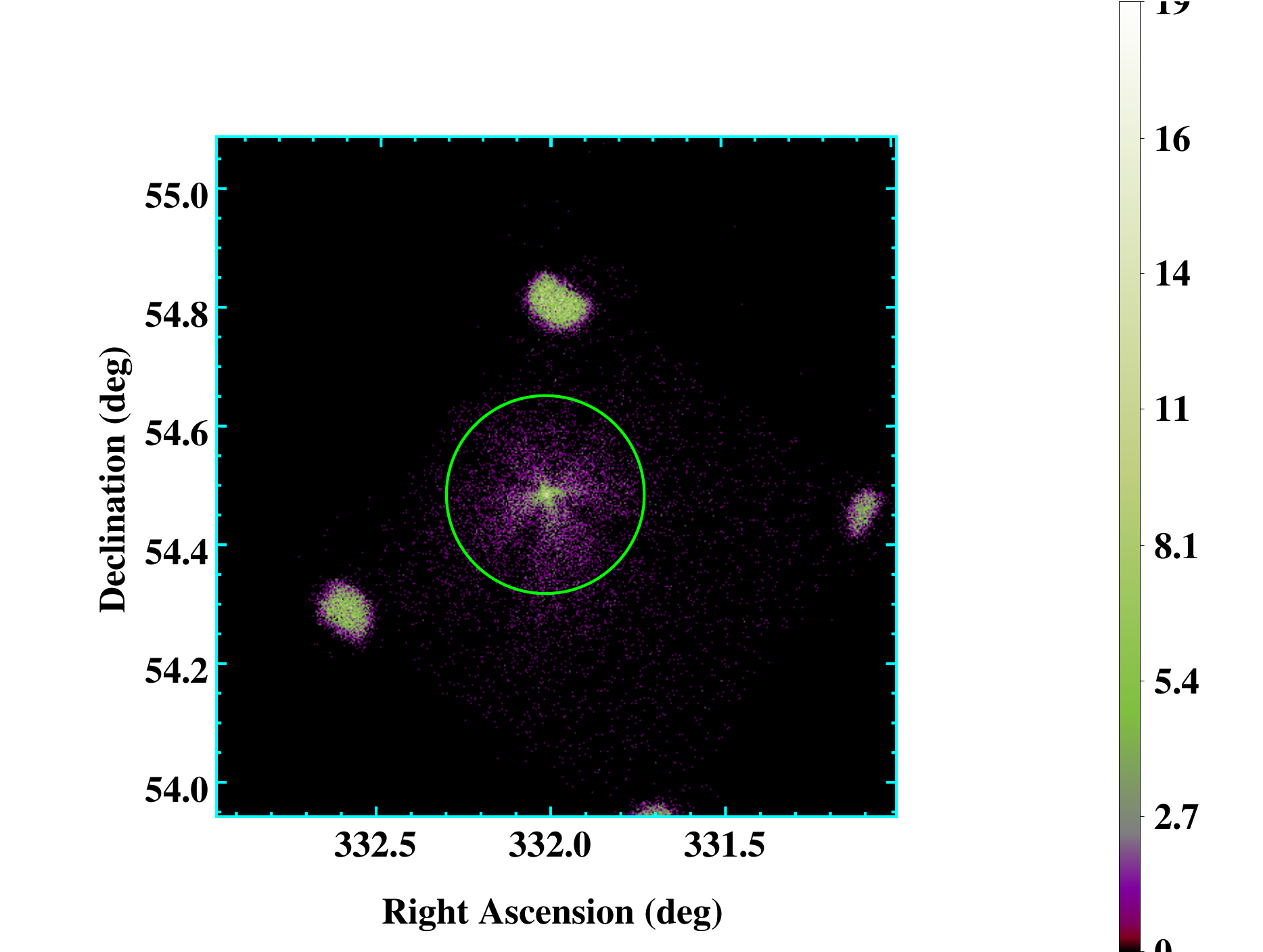}
\caption{The soft X-ray image of the \fourU in 0.3-7 \kev within \astrosat/\sxt FOV is shown. A circular region of 10\arcmin\ centered around the source is shown in green.} 
\label{fig:2}
\end{figure}

\subsubsection{\sxt \ Observations}
\label{sxt-obs}
The SXT observations of \fourU were carried out as a Regular Pointings in Photon Counting (PC) mode  with a total on 
source effective exposure of $\sim26.1~ks$ during the \astrosat orbits 5574 (MJD 57669.67) to 5589 (MJD 57670.69) 
(excluding the orbit-wise overlap of 3.9~ks).
We have obtained the SXT level-2 data of our source from {\it ISRO Science Data Archive for Astrosat Mission}
\footnote{\url{https://astrobrowse.issdc.gov.in/astro_archive/archive/Home.jsp}} located at 
ISSDC\footnote{\url{https://www.issdc.gov.in/astro.html}}. These were processed from the level-1 data 
with the SXTPIPELINE version 1.4a \footnote{\url{https://www.tifr.res.in/astrosat\_sxt/sxtpipeline.html}}.

At first we have merged the available orbit wise SXT level-2 clean event data of our source using SXT Event Merger Tool. \footnote{\url{http://www.tifr.res.in/~astrosat_sxt/dataanalysis.html}} 
This resulted in a single clean event file with an effective exposure of $\sim21.4$\ksec (from ObsID: 900000644)
and $\sim26.1$\ksec (from ObsID: 900000720). We have used XSELECT to extract an image of the source from the cleaned 
event files within the SXT field of view. We have considered a $10\arcmin$ circular region and an annular region of 
inner and outer radius of $13\arcmin$ \& $15\arcmin$ centered at the source as the background region to obtain the  source and 
background products (spectra, lightcurves) respectively from the \sxt observations.
We have obtained the background subtracted source light curves of \fourU 
 in three different energy ranges (0.3--3\kev, 3--5\kev \& 5--8\kev) with the appropriate range of channel selection within 
 {\sc XSELECT}. The low energy (0.3--8\kev) \sxt spectra of \fourU has been obtained from the SXT level-2 merged clean event file.
 A brief description of obtaining SXT spectra of \fourU and its subsequent analysis has been explained  in section~\ref{sxt-spectrum}.

\subsubsection{\laxpc \ Observations} \label{lxp-obs}
In the present paper, we have used the \laxpc observations of \fourU carried out in Event Analysis (EA) mode for our analysis 
and investigating the spectral and temporal variations of source. The level-1 data from all the three \laxpc units corresponding 
to EA mode observations of \fourU (i.e having \astrosat OBSID:9000000720 have been obtained from the \astrosat data archive at 
ISSDC. 
In order to follow up the recent spin period changes in \fourU (as described in the section~\ref{pulse-period}), 
we have also used another archival \astrosat/LAXPC observation of \fourU (Observation-ID : 9000000644). This observation has been 
carried out near the MJD 57637 with an effective exposure of $\sim43.7~ks$. A log of these observations is given in Table~\ref{table:1}.
The {\astrosat Science Support Cell}~\footnote{\url{http://astrosat-ssc.iucaa.in/?q=laxpcData}} \citep{2021JApA...42...28R} 
provides various software packages related to \laxpc observations to generate level-2 event data and subsequent extraction 
of source spectra and light curves. Among these we have used the LAXPCsoftware (August 15,2022 version) to extract the science 
products from our observations. 

\subsubsection{\czti Observations}
In our study, the observations of \fourU  taken with \czti in its Normal mode (M0) of operation are 
considered. A log of \czti observations are shown in Table~\ref{table:1}. The individual events recorded in this mode have a timing accuracy of 20\musec. For the analysis of \czti data, we have used the {\it cztpipeline} (version 3.0) \footnote{\url{http://astrosat-ssc.iucaa.in/cztiData}}.
We have followed the instructions as described in the CZTI analysis software user guide 
\footnote{\url{http://astrosat-ssc.iucaa.in/uploads/czti/CZTI_analysis_software_userguide_v3.0.pdf}} 
to process the "Level 1" data into "Level 2" standard data products. Using the {\it cztpipeline} 
module of the analysis software, we have generated the spectrum, spectral response and light curves 
in different energy bands for both the \astrosat observations of \fourU. The coded mask imager in \czti also measures the background simultaneously. Using this the background-subtracted products were generated. We have used the the module {\it cztbindata} of the pipeline with the mask-weighting technique to produce background subtracted source light curves and spectrum.

\subsection{\hxmt Observations}
Launched in June 2017, the Hard X-ray Modulation Telescope (a.k.a \hxmt)\citep{2020Ap&SS.365..158Z}
consists of 3 main scientific payloads. These are Low-Energy X-ray Telescope(LE, 0.7-13 \kev, 
 \citealt{2020SCPMA..6349505C}), the Medium Energy X-ray Telescope (ME, 5-30 \kev, \citealt{2020SCPMA..6349504C}) 
 and the High-Energy X-ray Telescope (HE, 20-250 \kev, \citealt{2020SCPMA..6349503L}) for scanning and pointed observations.   
 
\fourU has been observed with \hxmt at two epochs during 10-11, October 2020 with observation IDs: P0305070003 and P0305070004. A log of these observations along with the effective exposures in each telescope is shown in Table~\ref{table:1}. For the analysis of these observations, we have used HXMT 
 Data Analysis Software Package {\sc hxmtdas v2.04} and the calibration database 
 {\sc caldb v2.06} in our work. We have used 
 the science data taken only with small FOV  detectors from HE, ME and LE telescopes. 
 For the data calibration, screening and to generate  the high level data 
 products such as spectra, light curves and response and background files, we have followed the instructions as mentioned in the HXMT Data Reduction Guide (v2.04)
 \footnote{\url{http://hxmtweb.ihep.ac.cn/SoftDoc/496.jhtml}}. A barycentric correction is applied with the tool {\it hxbary} to the screened event files from all the detectors before extracting the lightcurves/spectra.     
\begin{figure*}
 \begin{center}   
    \includegraphics[height=8 cm,angle=-90]{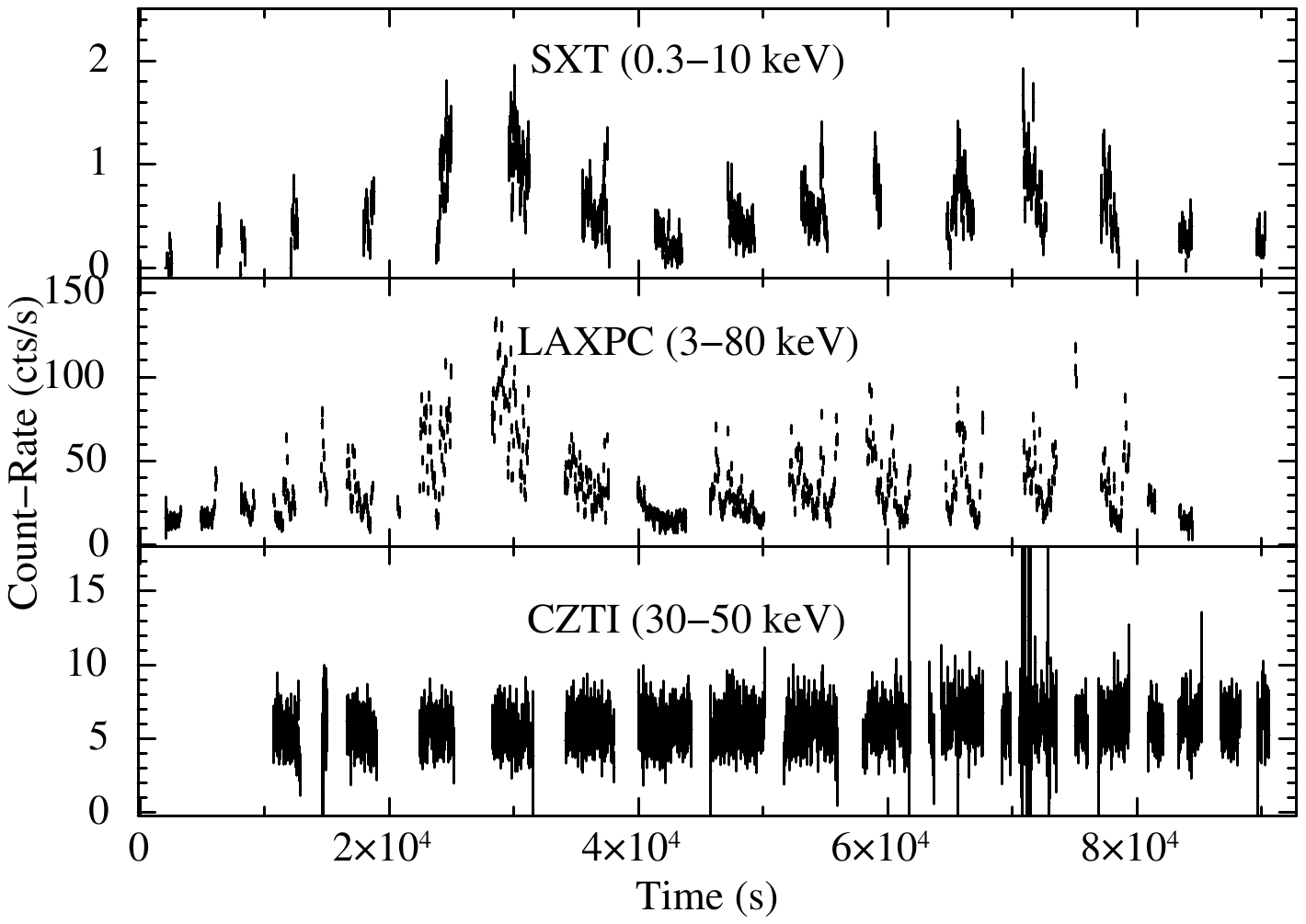}
    \includegraphics[height=8 cm,angle=-90]{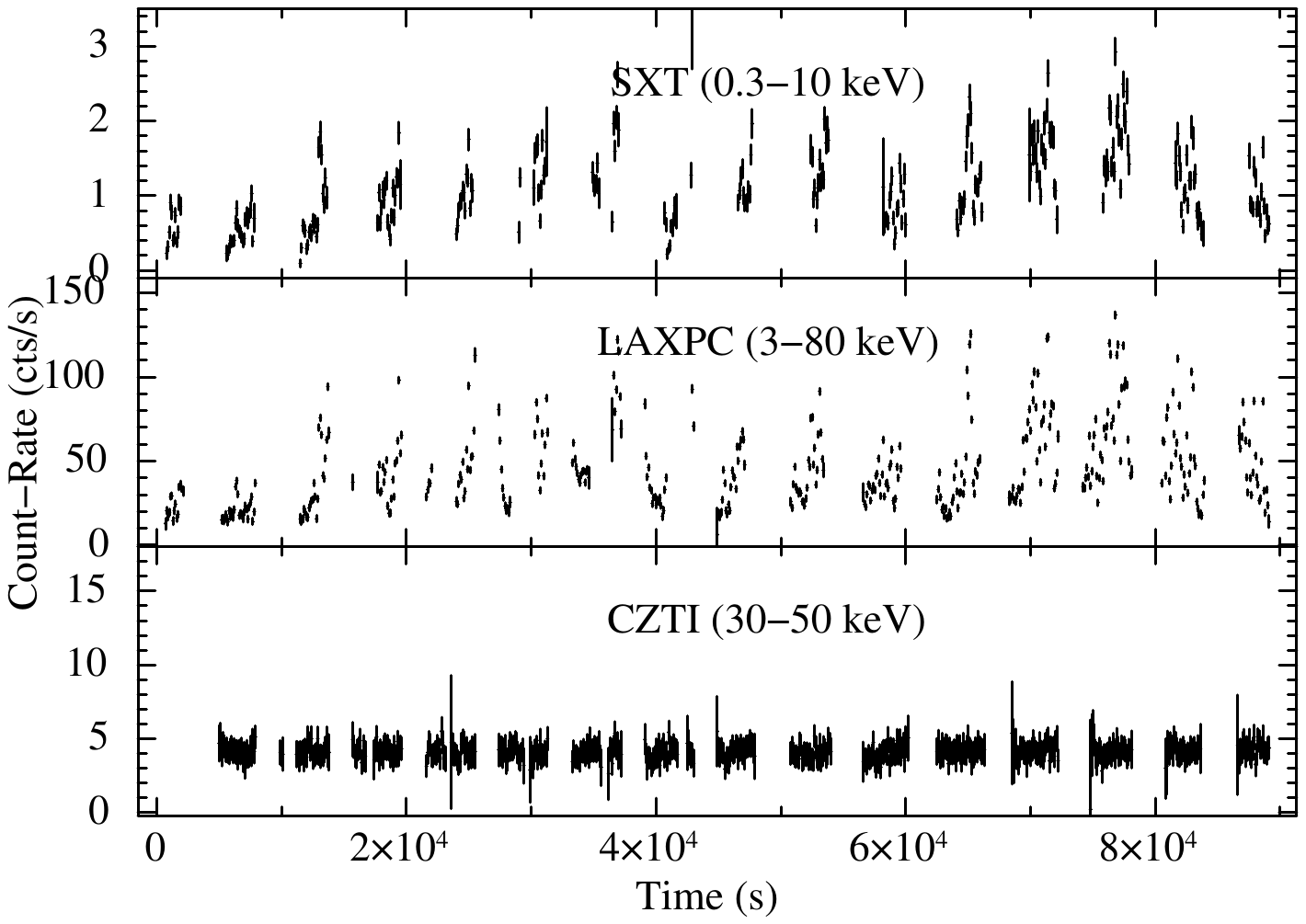} \\
    \includegraphics[height=8 cm,angle=-90]{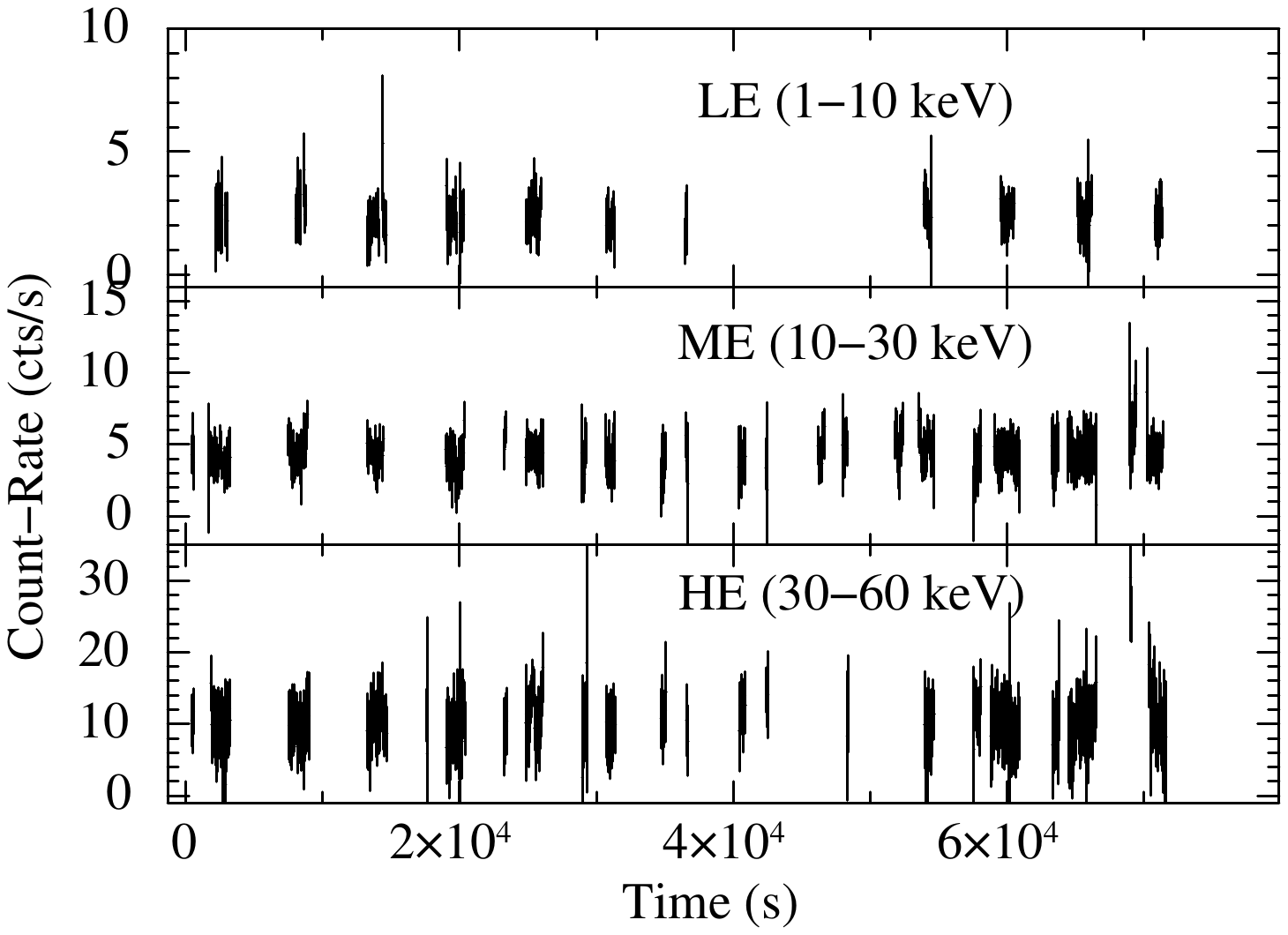} 
    \includegraphics[height=8 cm,angle=-90]{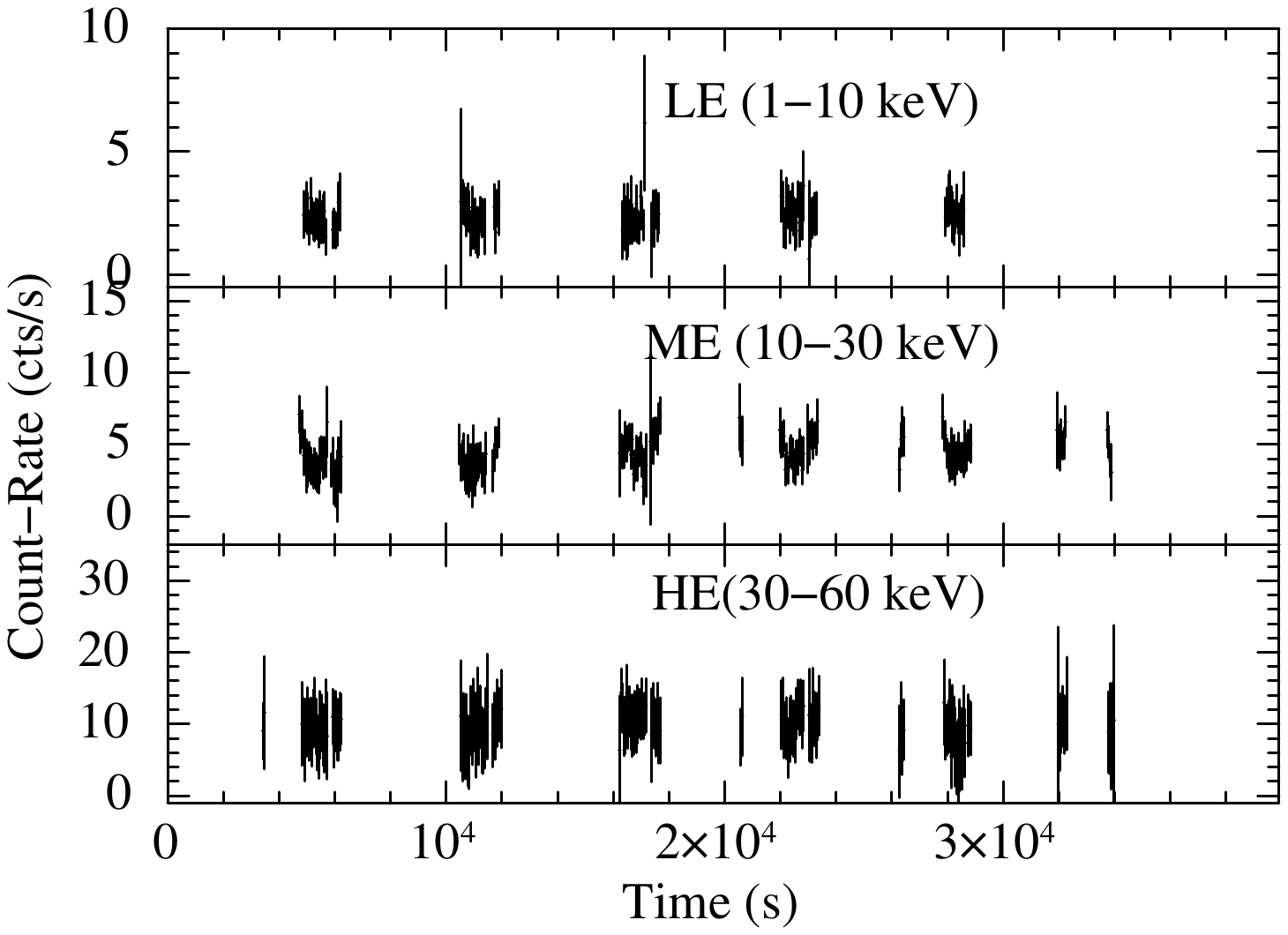} \\
    \caption{The light-curves of \fourU in different energy bands obtained from \astrosat \& \hxmt observations. 
     The light-curves obtained from \astrosat observations with OBSIDs: 9000000644 (top left), 9000000720 (top 
     right) and from the \hxmt observations with OBSIDs: P0305070003 (bottom left), P0305070004 (bottm right) are 
     shown in the figure. The insets in each figure represent the X-ray energy-bands and the  corresponding 
     instruments from which the background corrected light-curves (mask weighted correction in case of \czti) 
     are obtained. The light curves are binned with a 100 sec time bin.}
    \label{fig:3}
\end{center}    
\end{figure*} 
%
\section{Timing analysis} \label{sec:timing}
In order to study the timing variability of the long period pulsar in \fourU and to estimate its 
spin period pertaining to the present observation epoch, we have extracted the source and background 
lightcurves from the respective level-2 event files of SXT \& LAXPC observation. With the help a 
GTI file created (with the screening criteria as mentioned in section~\ref{lxp-obs}) for the 
\laxpc observations, we have extracted the source light curve in 3 -- 80\kev  with a binning of 
5 sec from the \laxpc-10 unit considering all the layers. Similarly, using the script 
``laxpc\_make\_backlightcurve'' provided with LAXPCsoftware, the background lightcurves were generated 
for the same energy range. We performed  barycentric correction to the background 
subtracted source light curve using the task {\it as1bary}. This barycenterd corrected 
light curves of \fourU obtained from \sxt, \laxpc-1, \czti respectively in 0.3--10\kev, 
3--80\kev and 30--50\kev are shown in Fig.~\ref{fig:3} for both the \astrosat observations. 

\begin{figure}
 	\begin{center}
            \includegraphics[width=0.98\linewidth]{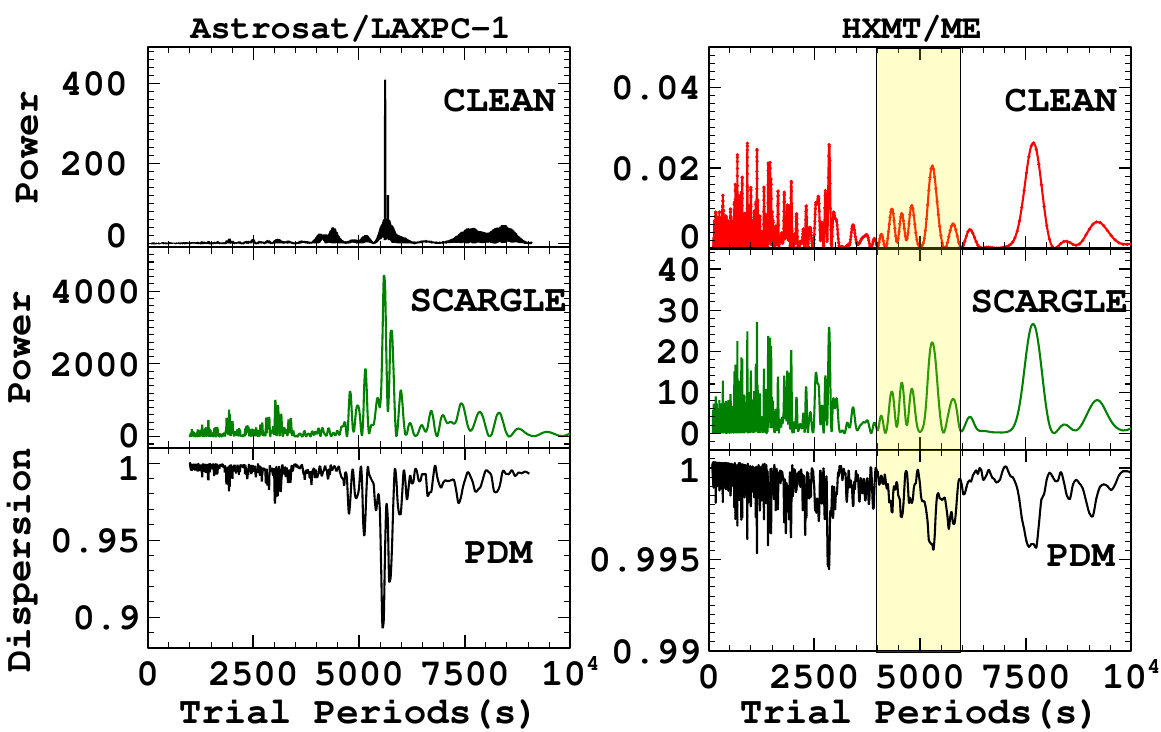}
 	\end{center}
 \caption{Periodograms from the lightcurves of \fourU obtained using the CLEAN, SCARGLE and the PDM algorithms. 
 The periodograms on the left are obtained from the combined \astrosat/LAXPC-1 observations 
 (ObsIDs: 9000000644, 9000000720) in 3-80 keV. The periodograms on the right are obtained from the  combined \hxmt/ME 
 observations (ObsIDs: P030507003, P030507004) in 10-30 keV. From these the estimated values of spin periods are shown 
 in Table~\ref{starlink}.} 
 \label{fig:4}
 \end{figure}

From the \hxmt observations, the background corrected and barycentred light curves of \fourU obtained 
in different energy  ranges as shown in Figure(s):\ref{fig:3} and \ref{fig:6} for both the 
observations. Since each of the \hxmt observations were segmented into exposure-IDs, we have used the {\sc FTOOLS} task {\it ftmerge} 
to combine the light curves from both the observations in a given energy range from each exposure-IDs into a single light curve. Similarly, we have used the combined the ME light curves in 10-30 \kev obtained from both the \hxmt observations to obtain the periodograms as shown in Figure~\ref{fig:4}. For the timing studies of \fourU with \astrosat, we have followed similar procedure. We have combined 
the light curves obtained at similar energies of \laxpc-1 from both the \astrosat observations to generate a single light curve. Similarly, the light curves are obtained from \sxt and \czti observed energy ranges. These light curves are further used to search for pulsations and generate pulse profiles in different energy range. 
%
\subsection {Determination of NS spin period}
\label{pulse-period}
The $\sim5500~s$ long spin-period of the pulsar in \fourU has previously been known to show long-term spin down 
\citep{2012MNRAS.425..595R,2013MNRAS.432..954W,2018MNRAS.479.3366T}. In order to determine the present value of 
its NS spin period in \fourU and its possible trend, we have extracted a light curve  in 5 -- 10 \kev from the 
LAXPC-1 observation. From this a background subtracted and barycentric corrected source light curve has been 
used to estimate the pulsar spin-period. The \astrosat/LAXPC observations of \fourU span a total duration of 
88.4~\ksec of which the effective pointed observation on source is $\sim40.7$~\ksec after applying the GTI selection. 
We have shown the 5--10 \kev light curve of \fourU in the inset of Fig.\ref{fig:3} which is also binned with 
$100\sec$ time bin. It can be seen from the figure that, the observed data is sparse and discontinuous containing 
data gaps to avoid the SAA Anomaly. This is also the case with \nustar observations of \fourU \citep{2018MNRAS.479.3366T}. 
 
Due to the long spin period of the pulsar, it can be seen  from the inset of Fig.~\ref{fig:3} that, the \astrosat 
observations also have incomplete phase coverage and thus restricting the use of any Fourier techniques for an ab-initio 
search of periodicity in the light curve. Secondly after applying the GTI selection to the LAXPC observation, the 
individual data segment within the LAXPC observation have lesser duration than the spin period of the pulsar. 
Therefore we have used period finding methods like \textsc{CLEAN} 
\citep{1987AJ.....93..968R}, \textsc{SCARGLE} \citep{1982ApJ...263..835S, 1989ApJ...338..277P}, the phase dispersion minimization 
technique (\textsc{PDM}; \citealt{1978ApJ...224..953S}) \& the classical discrete Fourier transform (\textsc{FT}; 
\citealt{1975Ap&SS..36..137D}). These algorithms were a part of {\sc PERIOD} (version 5.0), a time-series analysis package  
 of the {\sc starlink}\footnote{\url{http://starlink.eao.hawaii.edu/starlink}} software package \citep{2014ASPC..485..391C}. 
With the application of discrete Fourier transform method of Period04 \citep{2005CoAst.146...53L} program, we get a periodic signal 
at frequency $\nu=0.0001768967~Hz$ (from \astrosat) and at $\nu=0.0001885529~Hz$ (from \hxmt) observed light curves. The error on 
the frequency estimate is obtained within the Period04 program, wherein we have used the results from 500 Monte-Carlo simulations
of the processes. Each process consists of creating a time string data based on the original time string data of the light curve. 
The simulated count rate is then predicted by the last fit plus Gaussian noise.
 
\begin{figure*}
 	\begin{center}
 		\includegraphics[height=7 cm,angle=0]{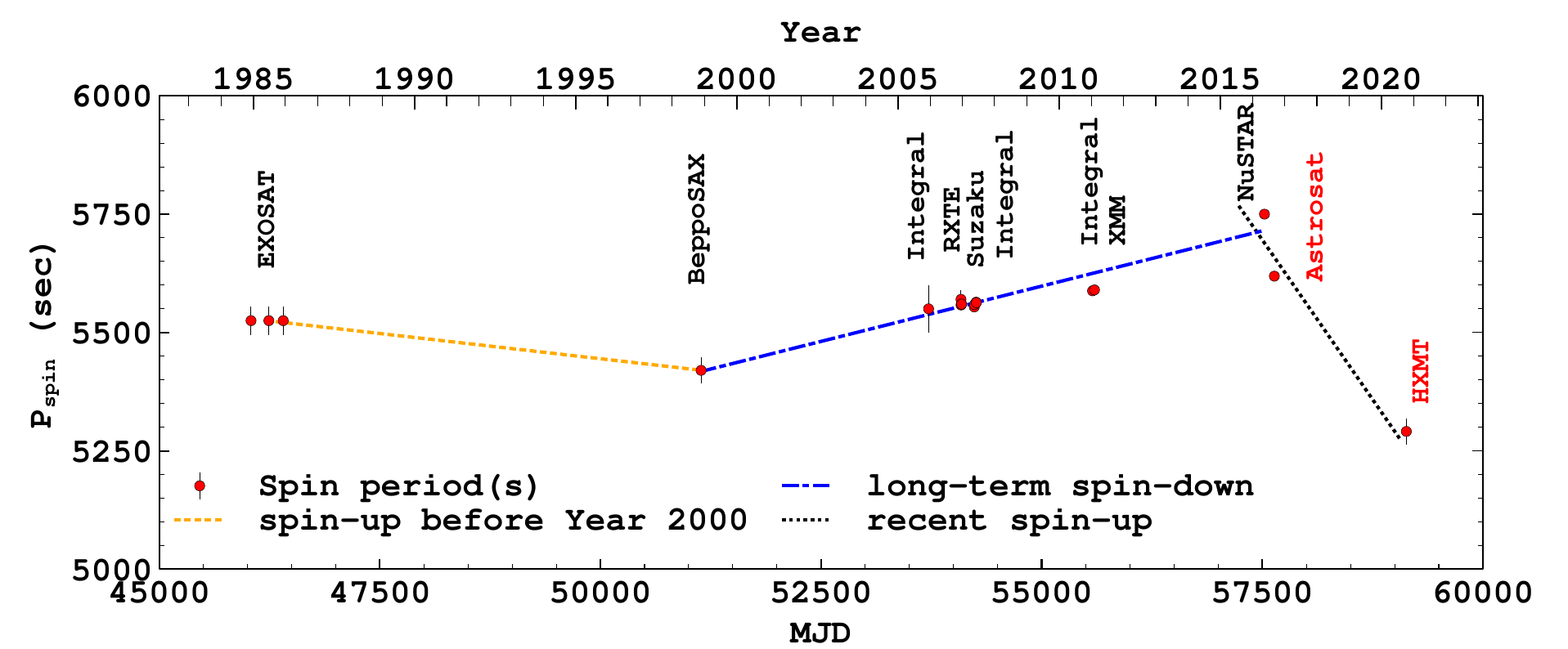}
 	\end{center}
 \caption{Long term spin period evolution of pulsar in \fourU. The linear fit for the recent 
 spin-up episode is obtained from \nustar\ to \hxmt observation of \fourU as: 
 $P_{spin}$=$21202.8-0.2697*MJD$ with an estimated pulsar spin-up rate of 
 $\left(1.2\pm0.1\right)\times10^{-13} Hz~s^{-1}$.} 
 \label{fig:5}
 \end{figure*}

In Table~\ref{starlink}, we show the spin-period estimates of \fourU obtained from these procedures. We have also performed a significance analysis of the peaks obtained with these procedures. The calculation of false alarm probabilities lie between 0.00 and 0.01 implying the peaks are significant with a 95\% confidence.   
We have considered the spin period value obtained from \textsc{CLEAN} method as our best estimate. A comparison of recent NS spin period evolution 
in \fourU from \astrosat \& \hxmt observations over its past 30 years evolution is shown in Fig.~\ref{fig:5}. The spin period 
obtained from the \astrosat  observations shows that the pulsar is spinning up recently after reaching its slowest period of 
$\sim5750~\sec$ as seen from \nustar observations \citep{2018MNRAS.479.3366T}. 
We note that, previously \citet{2022JApA...43..101J} found the spin period to be $5648(4)$~s from the second \astrosat observation (OBSID: 9000000720) using the Lomb-Scargle periodogram method. Whereas the use of \chisq maximization technique led them to a spin-period estimate of $5608(4)$~s for the same observation. It can be seen that, with the use of different period finding methods, the spin-period estimates are found to varying between 5608 s to 5652 s. This could be due to the presence of data gaps in the  observed light curves (see Fig~\ref{fig:3}) leading to incomplete phase coverage of the observed signal. Secondly, this also may suppress the power of the fundamental peak in a time series data (see \cite{2018MNRAS.479.3366T}). This can be seen from the periodograms obtained from HXMT/ME observed lightcurves (Figure~\ref{fig:4}).   
Our estimate of spin period  from \hxmt 
observations of \fourU further confirm the rapid spin-up of the pulsar (as can be seen in Fig.\ref{fig:5}). The latest spin-down 
rate of the pulsar until \nustar observation was $\left(-1.8\pm0.1\right)\times10^{-14} Hz~s^{-1}$ \citep{2018MNRAS.479.3366T}.
The subsequent spin-up episode of the pulsar in \fourU as observed with \astrosat and \hxmt observations in the present work 
indicates a net spin-up rate of $\left(1.2\pm0.1\right)\times10^{-13} Hz~s^{-1}$. This is an order of magnitude higher than the 
long-term spin down trend of the pulsar. The previously reported spin-down rate of the pulsar were
$\dot{\nu}$ = $\left(-1.7\pm0.3\right)\times10^{-14} Hz~s^{-1} $ \citep{2010ApJ...709.1249F} and
$\dot{\nu}$ = $\left(-1.5\pm0.2\right)\times10^{-14} Hz~s^{-1}$ \citep{2012ASPC..466...25R}.
 
\begin{table}
	\begin{center}
		\caption{Estimation of spin period(s) of \fourU from \astrosat and \hxmt observations.}
		\label{starlink}
		\begin{tabular}{lcc}
			\hline 
            Instrument  &Astrosat/LAXPC    &HXMT/ME \\
			OBSID       &(644+720) &(7003+7004) \\ 
   			\hline \noalign{\smallskip}
			Method		&$P_{\rm spin}$ (s) &$P_{\rm spin}$ (s)	\\
			\hline \noalign{\smallskip}
			CLEAN	&$5619\pm3$ &$5291\pm28$	\\
			SCARGLE	&$5653\pm3$ &$5291\pm28$	\\
			PDM		&$5627\pm3$ &$5319\pm29$	\\
			CHISQ   &$5616\pm3$ &$5291\pm28$    \\
			DFT (period04) &$5652\pm2$   &$5303\pm17$        \\
			\hline
			Mean		&$5633\pm2$  &$5299\pm5$	\\
			Weighted mean	&$5637\pm1$ &$5300\pm11$	\\
			\hline
			Final adopted value	&$5619\pm3$ &$5291\pm28$ \\
			\hline
		\end{tabular}
	\end{center}
\end{table} 

\subsection {Pulse profiles and pulse fractions}
\begin{figure}
 \begin{center}
 	\includegraphics[height=8cm,angle=0]{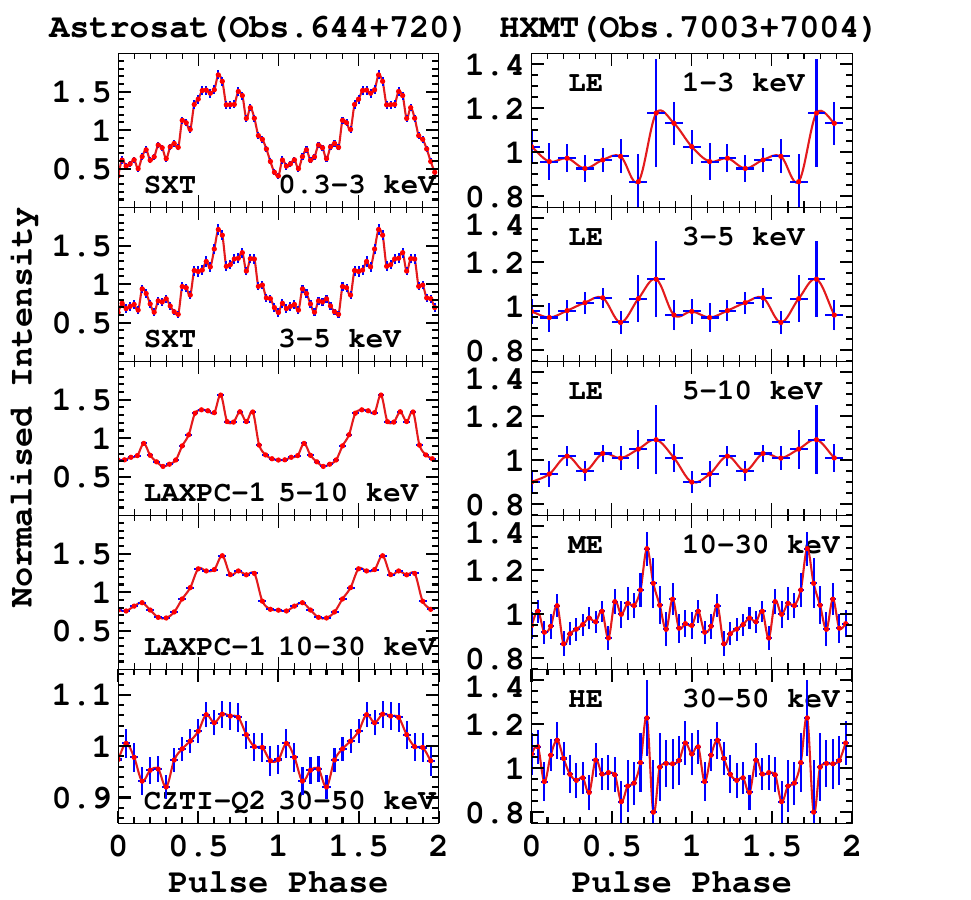}
 \end{center}
 \caption{Comparing the energy resolved pulse profiles of \fourU  obtained from \astrosat 
  (panels in left) and \hxmt (panels in the right) observations.  The pulse profiles were 
  obtained with a folded period of $5619\sec$ at MJD 57637.076 with  light curves obtained 
  from \astrosat observations. And with a folded period of $5291\sec$ at MJD  59132.545 to 
  the light curves obtained from \hxmt observations in the respective energy bands as shown 
  in the inset of each plot. Two complete pulse profiles are shown in each panel for the clarity.} 
 \label{fig:6}
 \end{figure}

\begin{figure}
 	\begin{center}
 		\includegraphics[height=5.5cm,angle=0]{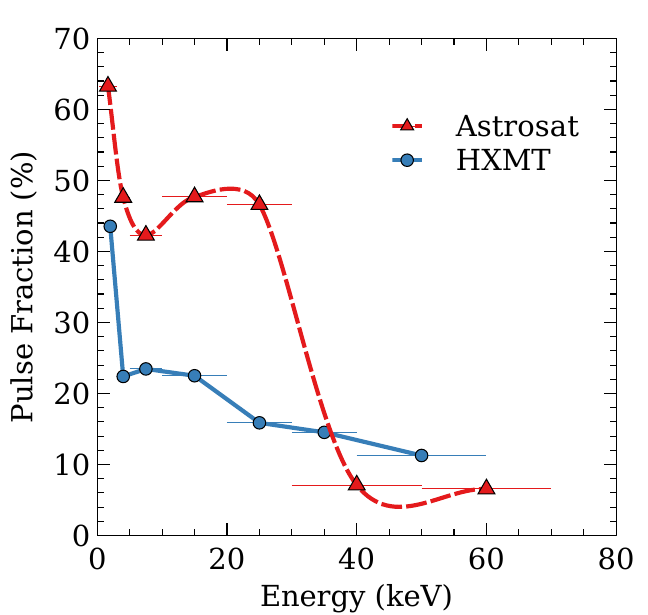}
 	\end{center}
    \caption{Pulse Fraction changes with energy as seen during \astrosat \& \hxmt observations of \fourU.}
    \label{fig:pf}
    \end{figure}

We have folded the lightcurves obtained from combined \astrosat observation with the  estimated spin period of $5619\pm3$~s with the epoch of folding at MJD 57637.076 
to obtain the pulse profiles in different energy range. In Fig.~(\ref{fig:6}) these pulse profiles are shown in $0.3-50$~\kev. 
The pulse profiles are broadly single peaked with the appearance of narrow structures in the 0.5 -- 0.7 
phase bins in 0.3--5 \kev. The shape and structure of the profile remains similar across different energy bands. 
There are no major energy dependence features seen in the pulse profiles upto 30\kev during \astrosat observations. 
The shape of the pulse profile remains same with energy upto 50\kev. At hard X-rays energies above 30\kev 
the pulse profiles become smooth and single peaked with the disappearance  of narrow structures as seen at 
lower energies during the \astrosat observations. The pulse-profiles from \hxmt observations are obtained from folding the 
HE, ME and LE lightcurves with the spin-period of $5291\pm28$~s considering the folded epoch at MJD 59132.545.
These pulse profiles in different energy ranges seen with \hxmt show marginal pulsations (Fig.\ref{fig:6}).
A prominent peak at 0.7--0.8 pulse phase bins is clearly seen at 10-30 \kev. Beyond which, it starts diminishing 
and comparable  with the other structures (Fig.~\ref{fig:pf}).  Unlike \astrosat observations, we do not see clear pulsations 
in 30--50\kev and beyond during \hxmt observations of \fourU.
 
The average pulse-fraction (PF defined as the ratio of intensity at the maximum minus the minimum of
pulse profile to the maximum plus minimum intensity ) is found to be $63\%$  in soft X-rays in 0.3-3 \kev. Thereby a decreasing trend upto $\sim42\%$ in 5-10 \kev. There is a marginal increase of pulse-fraction value to $47\%$ in 10-20 \kev. Subsequently, there is a decreasing trend seen up to 30\kev with a further decline up to 7\% in 50-80 \kev. Beyond 30\kev the pulse-fraction is marginal (\ie$<10\%$) in hard X-rays.  The changes in PF value with energy during \astrosat \& \hxmt observations is shown in Figure~\ref{fig:pf}.
  
\section{Spectral analysis} \label{sec:spectra}
The broad-band continuum spectra of \fourU in 1--60~\kev obtained from  \astrosat observations  is shown in Fig.~\ref{highecut1}. For this we have used the spectra obtained from  \sxt, \laxpc 
and \czti.The spectra of \fourU obtained from \hxmt observations were found to be not suitable 
for spectral studies, because the spectra are mostly background dominated.


\subsection{The SXT spectrum}\label{sxt-spectrum}
The soft X-ray spectrum of the source from SXT observation is obtained from the merged clean 
event file (see section~\ref{sxt-obs}). For the extraction of the source spectrum we have used 
the {\sc XSELECT} with the same region filter (i.e 10$\arcmin$ around the source with coordinates 
centered at RA: 332.0162, DEC: 54.4884) as that was used earlier to extract source light curves 
in different energy ranges. For the PC mode observations of our source, the required  SXT background 
spectrum file and the instrument response file as provided by the SXT instrument team has been used in our spectral analysis. An off-axis ancillary response file (ARF) that is needed during the spectral fitting has been created using the SXT Off-axis ARF Making Tool (sxtmkarf) from an input ARF file 
\footnote{{\url{http://www.tifr.res.in/~astrosat\_sxt/dataanalysis.html}}}. 
For the use of SXT source spectra in spectral fitting, we have used the data in $1.0-7.0$~\kev while applying a binning scheme of having a minimum of 30 counts per bin to use $\chi^{2}$-statistics in spectral fitting. 


\subsection{The LAXPC spectrum \label{laxpc-spectra}}
We have obtained the source and background spectra from \laxpc units as briefly mentioned in the section~\ref{lxp-obs}. The source spectra from the LAXPC-20 \& LAXPC-30 units in the energy range 
of 4-30\kev has been used in our spectral analysis. We have restricted the \laxpc observed source spectra upto 30\kev beyond which, it is mostly the X-ray background. The response files of these detector units close to the epoch 
of observations were used from the LAXPC Response files provided by Astrosat Science Support Cell (ASSC) \footnote{\url{ http://astrosat-ssc.iucaa.in/?q=laxpcData}}.

\begin{table}
	\centering
    \caption {Best-fitting spectral parameters with 2$\sigma$ (90\%) errors obtained from \astrosat observations (ObsID: 9000000644, 9000000720) of \fourU . The 1-60\kev composite spectra is fitted with a empirical continuum model such as a power law model with a High-energy cutoff (\ie \hcp\ ).}
	\begin{tabular}{llll}
		\hline 
		Model             &Parameters        &9000000644      &9000000720    \\ 
		\hline
	      \tbabs	        &N$_{H1}$~($(10^{22}$~\pcm)     &3.1$\pm$0.2     &1.7$\pm$0.1\\
		  \tbpcf            &N$_{H2}$~($(10^{22}$~\pcm)     &11.4$\pm$1.4    &-        \\ 
                              & Cvr. Frac.                    &0.66$\pm$0.04   &-    \\
		  \pl               &$\Gamma$                       &1.6$\pm$0.1     &1.2$\pm$0.1  \\
                              &$Norm_{PL}$       & $0.073\pm0.014$ & $0.017\pm0.002$ \\
		  \hcp              &$E_{cut}$ (keV)	            &7.8$\pm$0.6     &4.1$\pm$0.4  \\
		                     &$E_{fold}$ (keV)               &18.7$\pm$2.2    &13.5$\pm$0.7 \\
            \\                  
		                     &$F^{\rm a}_{\rm 1-7\,keV}$     &3.45            &1.20\\
                              &$F^{\rm a}_{\rm 7-30\,keV}$    &1.86            &2.35\\
                              &$F^{\rm a}_{\rm 30-60\,keV}$   &0.67            &0.78\\ 
            \\                  
                              &$L_{x}^{\rm b}({\rm 1-60\,keV})$    &6.5             &4.5 \\

                              &\chiq (d.o.f)                  &1.01 (652)      &1.10(606) \\
		\hline
	\end{tabular}
	\flushleft
	Notes:
	$^a$ : Absorption corrected flux in unit of \flux; \\
	$^b$ : The 1--60 keV X-ray luminosity in the units of 10$^{35}$~\lum assuming a distance of 
               ($3.7\pm0.4$)~\kpc to the source.\\ 
	\label{table3}
\end{table}
\subsection{Phase-averaged spectroscopy}\label{section-spec}
The simultaneous observations of \fourU with \sxt, \laxpc and \czti instruments onboard \astrosat has allowed 
us to explore the broad-band spectra of the pulsar \fourU. In order to study its spectral properties, we have 
used the X-ray spectral fitting package \xspec(version 12.10.1) \citep{1996ASPC..101...17A} to simultaneously 
fit the \sxt, \laxpc and \czti spectra in the energy interval of 1-60~\kev. The spectra of \fourU has been fitted 
with various empirical models such as: a cutoff power-law model (\cpl), a high-energy cutoff powerlaw (\hcp), and
a two component cutoff power-law model having negative and positive exponential cutoff ({\sc NPEX})\citep{1995AAS...18710403M}. In our spectral
analysis we have used the above models, allowing a  $3\%$ systematic error in the fitting. These models are 
generally found useful in describing the spectra of accretion-powered X-ray pulsars in a broad luminosity range. 
To account for the galactic absorption of X-rays along the direction of the source, we have used the photoelectric 
absorption model component (\tbabs) along with the above empirical continuum models in our spectral fitting. 

We have seen that the above mentioned continuum models fit the spectra well. We have therefore chosen the \hcp\ model
explain the \astrosat spectra of \fourU. This model gives fits with better statistics with \chisq (dof) of 658.36 (652) 
and 669.7 (606) respectively for the first and second \astrosat observations (ObsID: 9000000644, 9000000720). 
The \cpl\ model applied to the phase-averaged spectra, results in comparable \chisq (dof) values of 639.4(627) and 
612.5(582) respectively for the above \astrosat observations. 
 \begin{figure*}
 \tabcolsep=0.01\linewidth
   \divide\tabcolsep by 3
   \begin{tabular}{cc}
 		\includegraphics[width=0.31\linewidth]{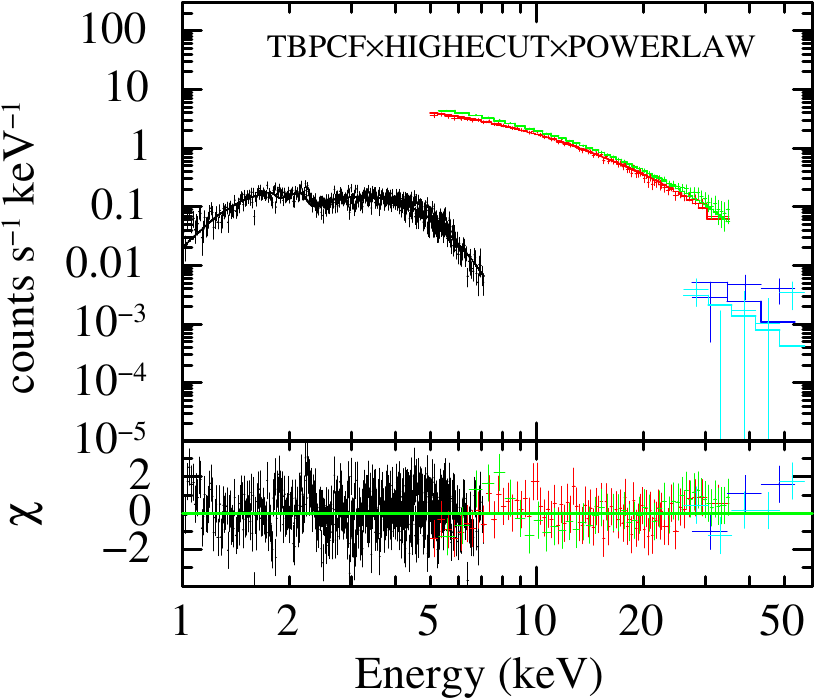} 
            \includegraphics[width=0.31\linewidth]{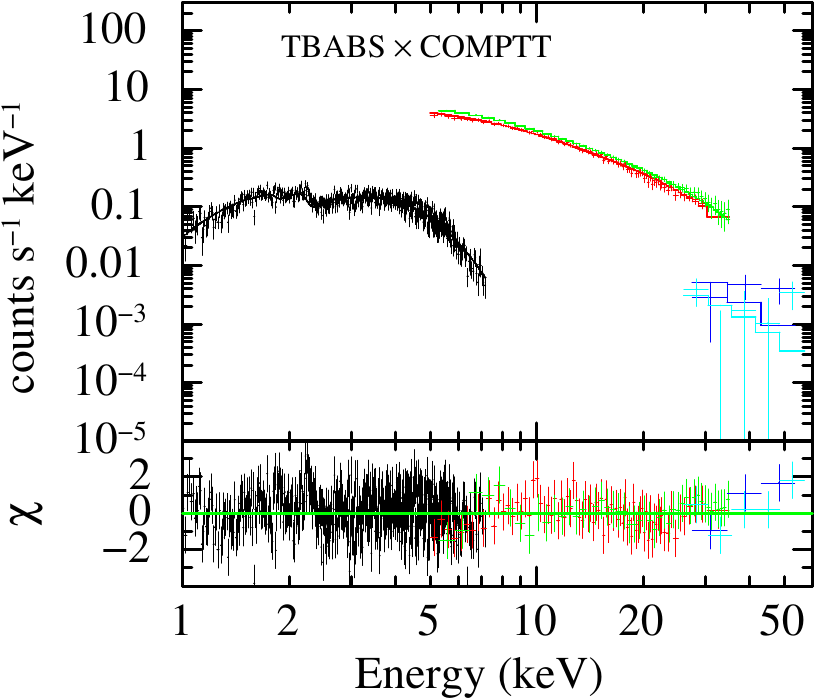} 
            \includegraphics[width=0.31\linewidth]{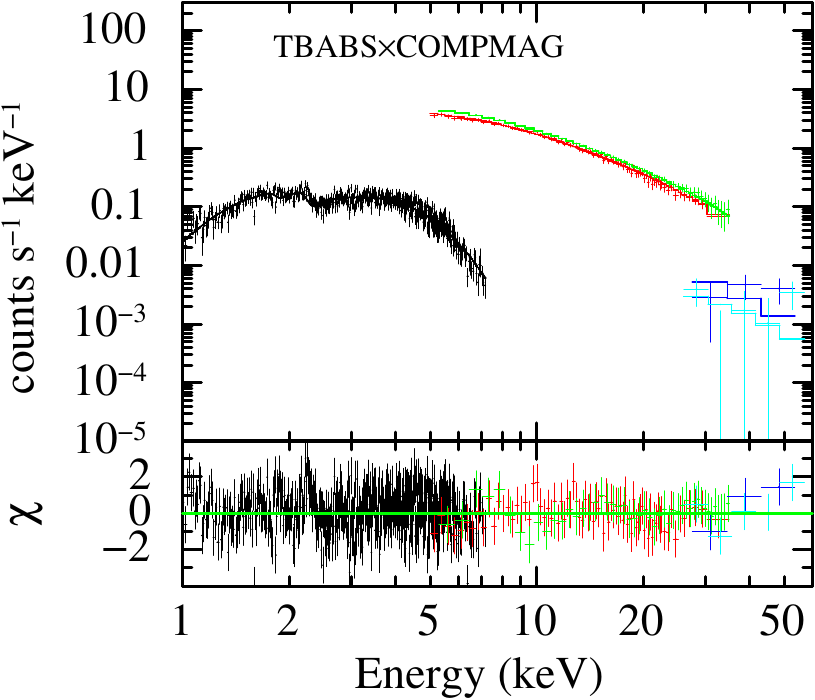} \\
            \includegraphics[width=0.31\linewidth]{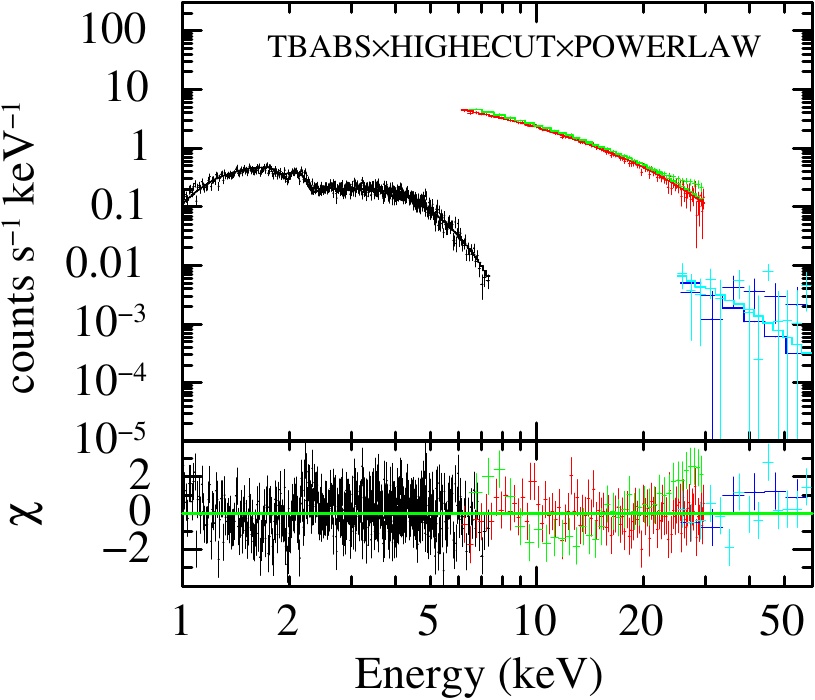}
            \includegraphics[width=0.31\linewidth]{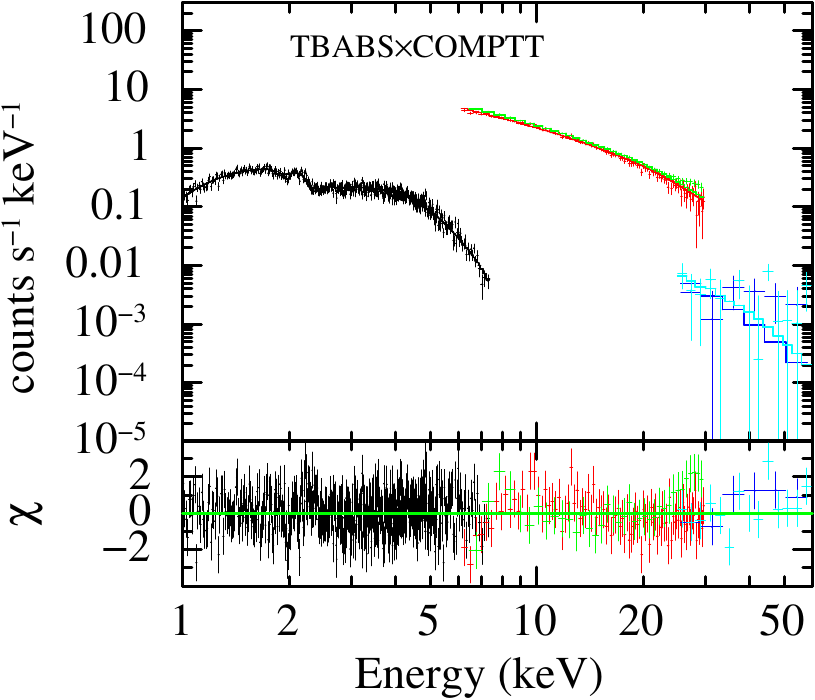} 
            \includegraphics[width=0.31\linewidth]{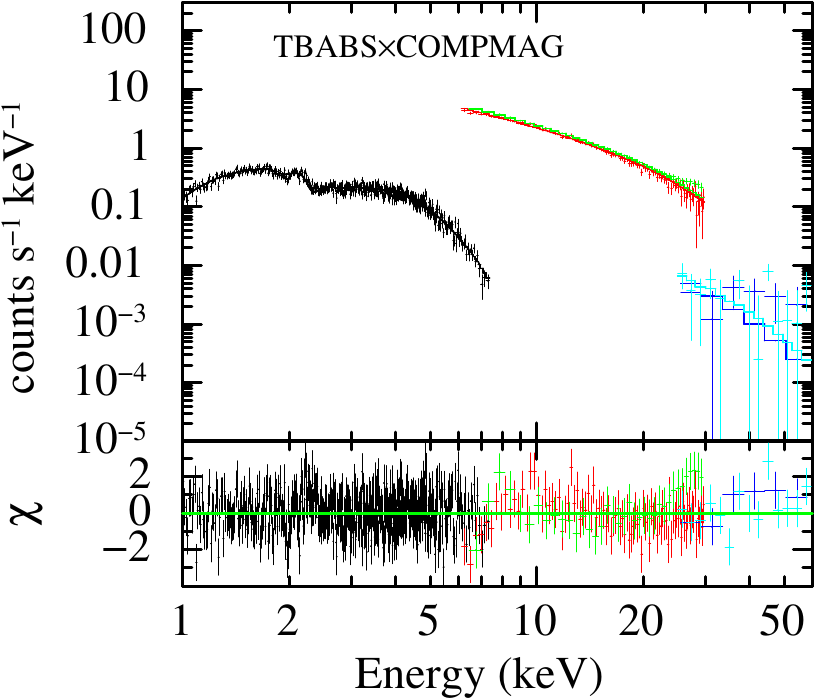} \\
  \end{tabular}
  \caption{The broadband phase averaged spectra of \fourU in 1--60\kev obtained from the \astrosat/\sxt(1--7\kev), 
  LAXPC(7--30\kev) and CZT(30--60\kev) from observations with OBSIDs: 9000000644 (figures in top panels) and 
  9000000720 (figures in bottom panels). The spectra is best fitted with models such as: an absorbed high-energy
  cutoff continuum model and Comptonization models such as: \compmag, \comptt. The second panels in each of the figure shows the spectral residual obtained after the best-fitted continuum model.}
 \label{highecut1}
 \end{figure*}

From the spectral residual, we find that the spectra from first \astrosat observation (\ie ObsID: 9000000644) 
require a partial absorbed component in addition to the galactic absorption component to better fit the spectra at 
soft X-rays. Whereas, for the second \astrosat observation, we do not require any partial absorbed component at 
low X-ray energies. The significance of additional absorbed component along with the high-energy Cutoff power law (\hcp) 
continuum model can be seen from comparing the values of chi square fit statistics and degrees of freedom  with and without 
the extra absorption component. We find that, with the inclusion of absorbed component (\tbpcf) results in a Chi-squared value 
of 658.36 with 652 degrees of freedom. Whereas without the absorbed component, the Chi squared  value of fit statistic is 
752.39 with 654  degrees of freedom (d.o.f). A ftest comparison of these two chi-square values yields  F-statistic value of
46.56 with a very low ftest probability of $1.255\times10^{-19}$. A similar need of additional absorbed component is also 
seen with the \cpl\ continuum model while fitting the broadband spectra of \astrosat observation (ObsID: 9000000644). 
In case of best fitted  \cpl\  model we have obtained the chi-squared value of 639.42 with 627 d.o.f  with the additional 
absorbed component and the chi-squared value of 772.86  for a 629 d.o.f. without the absorbed component. The subsequent 
ftest comparison of \cpl models for the inclusion of absorbed component yields a  very low Ftest probability of 
$1.56\times10^{-26}$ with a F statistic value of 65.42. Given the low values of ftest probability, it is reasonable to 
explain the broadband spectra with the additional absorbed component. 

We choose the \hcp\ model to explain the phase averaged spectra of both the \astrosat
observations in 1--60\kev. The best fitted spectral parameters obtained from the model are shown in Table~\ref{table3} 
along with the absorption corrected source X-ray flux estimates in soft and hard X-ray energy bands. The uncertainties 
in the spectral parameter estimates are quoted at  $1\sigma$ confidence level  for one parameter of interest. 
In the first panels of Fig.~\ref{highecut1}, we show the 1--60~\kev unfolded spectrum of the source along with the 
best-fitted high-energy cutoff continuum model. It can seen from Table~\ref{table3} that, the absorption corrected 
soft X-ray flux in 1--7\kev from  first \astrosat observation is $3.45\times10^{-10}$\flux comparatively high than that 
of the second observation by a factor of more than 2. Whereas the hard X-ray flux in 7--30\kev and 30--60\kev are of similar 
magnitude in both the \astrosat observations. Apart from these, the spectral parameter values such as: cutoff-energy $E_{cut}$, 
folding-energy $E_{fold}$,and the powerlaw spectral index ($\Gamma$) are found to be higher for the first \astrosat observation 
compared to the second observation. We note that, using only the \laxpc spectra in 3--30\kev of above \astrosat observations, 
\citet{2022JApA...43..101J} have explained the spectra with a \hcp\ model. Whereas, our broadband spectral 
coverage in 1--60\kev allowed us to identify additional absorption in soft X-rays of the first \astrosat observation.

\begin{table} 
\centering
\caption{Best-fit parameters of the Comptonization models \tbabs$\times$\comptt\  and \tbabs$\times$\compmag\ 
for \fourU broad-band spectra in 1-60~\kev obtained from \astrosat observations with observation-IDs 9000000644 
\& 9000000720  respectively. In case of \compmag\ model, we have fixed the model parameters $\eta$=1, $\b0$=0.05, $r_{\rm 0}$=0.25 and A=0.5 for both the \astrosat observations.}
\label{table:4}
\begin{tabular}{llll} 
\hline
 Model   &Parameter                      & 9000000644                & 9000000720  \\
\hline
\tbabs    &$N_{\rm H}$~($10^{22}$~\pcm)  & 1.62$^{+0.06}_{-0.06}$    & 0.82$^{+0.05}_{-0.05}$ \\
\compmag  &$\ktbb$ (keV)                 & 1.56$^{+0.03}_{-0.02}$    & 0.89$^{+0.02}_{-0.02}$ \\
          &$\kte$  (keV)                 & 14.9$^{+5.6}_{-1.0}$      & 7.1$^{+0.9}_{-0.5}$ \\
          &$\tau$                        & 0.22$^{+0.01}_{-0.04}$    & 0.36$^{+0.01}_{-0.01}$ \\
          &Norm                          & 28.7$^{+9.6}_{-5.3}$      & 57.8$^{+3.9}_{-3.8}$ \\
    \\
    &$F^{\rm a}_{\rm 1-7\,keV}$          & 1.77                     & 1.05\\
    &$F^{\rm a}_{\rm 7-30\,keV}$         & 1.8                      & 2.35\\
    &$F^{\rm a}_{\rm 30-60\,keV}$        & 0.78                     & 0.40\\
    &\chiq (d.o.f)                       & 0.98 (655)               &  1.02(606) \\
\hline \\
\tbabs  &$N_{\rm H}$~($(10^{22}$~\pcm)   & 1.01$^{+0.12}_{-0.11}$   & 0.59$^{+0.07}_{-0.06}$ \\
\comptt &$\kt0$ (keV)                    & 1.33$^{+0.05}_{-0.05}$   & 0.76$^{+0.02}_{-0.02}$ \\
        &$\kte$ (keV)                    & 8.05$^{+2.14}_{-1.08}$   & 6.79$^{+0.59}_{-0.46}$ \\
        &$\tau$                          & 3.06$^{+0.42}_{-0.56}$   & 3.63$^{+0.23}_{-0.24}$  \\
        &Norm                            &($6.9\pm0.7$)$\times10^{-3}$ &($4.5\pm0.1$)$\times10^{-3}$ \\
    \\
    &$F^{\rm a}_{\rm 1-7\,keV}$          & 1.7                     & 1.0\\
    &$F^{\rm a}_{\rm 7-30\,keV}$         & 1.8                     & 2.35\\
    &$F^{\rm a}_{\rm 30-60\,keV}$        & 0.64                    & 0.44\\
    &\chiq (d.o.f)                       &0.99 (668)               &1.02 (606) \\
\hline
\end{tabular}
   \begin{list}{}{}
    \item[$^{\mathrm{a}}$]{In units of $10^{-10}$ erg\,cm$^{-2}$\,s$^{-1}$}. 
   \end{list}
\end{table}
Secondly, we have seen that, only from the \laxpc data of first \astrosat observation, \citet{2022JApA...43..101J}
have reported a detection of broad emission line around 7\kev. However, from our spectral fitting of the same \laxpc 
observations along with \sxt observations in 1--7\kev and \czti observations in 30--60\kev, we do not observe any 
such line emission feature around 7\kev. In the second \astrosat observation also we do not find any significant 
line emission. The non-detection of any line emission in the broadband spectral fitting for the second observation 
agrees with the spectral results of \citet{2022JApA...43..101J}. Therefore we have not included any emission line 
component in the spectral fitting of \fourU spectra. In the hard X-rays, from \astrosat observed spectra, we do not 
see any signature of cyclotron line emission at 35\kev, that was anticipated earlier during \bepposax observations 
\citep{2004A&A...423..311M}. The non-detection of cyclotron line is also reported from  earlier from \integral 
observations \citep{2013MNRAS.432..954W} and recent \nustar\ observations of the pulsar \citep{2018MNRAS.479.3366T}. 

Apart from explaining the phase averaged spectra of \fourU with empirical continuum models, we have also 
studied the spectra with application of Comptonization models. 
Previous studies have shown that, the physical models based on Comptonization also provide good fits to the data 
\citep{2004A&A...423..311M,2009A&A...494.1073R,2018MNRAS.479.3366T}. Therefore we have also followed this approach 
to explain the \astrosat observed spectra of \fourU in 1-60\kev.
In our spectral fitting with these models, we find that any additional blackbody component is not needed at 
soft X-rays to explain broad phase averaged spectra. For both the \astrosat observed spectra of \fourU, we find
that the application of \comptt\ and \compmag\ Comptonization models could fit the spectra well. The \comptt model 
applied to first \astrosat observation (\ie ObsID: 9000000644) has a \chisq(dof) of 667.9(668) and for the 
second observation (\ie ObsID: 9000000720) has a \chisq(dof) of 617.9(606). The \compmag\ model similarly yields a 
\chisq(dof) of 642(655) and 616.5(606) for first and second observations respectively. To account for the galactic 
absorption along the line of sight, we have used \tbabs\ model with the ISM abudances and photoionization 
cross sections set to wilm abundances \citep{2000ApJ...542..914W}. We find that, the average H1 column density (NH) 
estimated through H14Pi map \citep{2016A&A...594A.116H}  for the sky position of \fourU is $0.55\times10^{22}$\pcm.
From the best fitted spectral parameters obtained from Comptonization models with \tbabs\ component as mentioned in 
Table~\ref{table:4},  we find that the estimated column density is similar to the value obtained from H14Pi map. 
For the first \astrosat observation, however, we see a slight increase of $N_{H}$ values albeit with higher values of blackbody
seed photon temperature ($kT_{0}$). The absorption corrected flux for the pulsar in 1--7\kev, 7--30\kev and 30--60\kev
obtained from \sxt, \laxpc and \czti observed spectra respectively are also estimated using best-fitted Comptonization models. 
The estimated flux values show similar values in both the Comptonization models. The net flux in 1-60\kev during first observation 
is higher compared to the second \astrosat observation of \fourU. 

The \compmag\ model \citep{2012A&A...538A..67F} assumes a cylindrical accretion geometry for the matter accreting onto 
the polar cap of a magnetized neutron star. Numerically, it solves the radiative transfer equations, assuming a cylindrical 
accretion onto a magnetized neutron star where both the thermal and bulk Comptonization of seed photons occur. The seed photon 
spectrum is primarily of blackbody in nature with an exponential distribution within the accretion column. 
\citet{2012A&A...538A..67F} further states that the \compmag\ model generally holds good for an optical depth $\tau\geq1$ 
as the model is based on Fokker-Planck approximation of the radiative transfer equation. In our best fitted spectra of \fourU 
with \compmag, however, we obtain $\tau$ values within a factor of 3 to unity. This has been explained in case of \xtej
\citep{2012MNRAS.424.2854F} that, such a scenario would arise for an inefficient diffusion of photons along the column axis. 
Therefore the photons may efficiently propagate perpendicular the column axis. The other possibility is that, emission being 
anisotropic due to strong magnetic field of the neutron star. 

The possibility of high magnetic field strength of magnetar nature \citep{2012MNRAS.425..595R,2013MNRAS.432..954W} further 
supports this in case of \fourU. Since the other physical parameter results such as electron temperature ($kT_{e}$) and 
blackbody photon temperature ($kT_{bb}$) are consistent within the estimated 1$\sigma$ errors with that from \comptt\ model 
parameter values (\ie $kT_{e}$ and $kT_{0}$), these variations can be discussed reasonably using \compmag\ physical model. 
From the Comptonization model parameters, we notice that, the soft photon source temperature ($kT_{bb}$) during the first 
\astrosat observation is substantially higher compared to thesecond \astrosat observation. The hotter blackbody temperature 
is observed in \fourU spectra of first \astrosat observation with both \compmag\ and \comptt\ model. In the spectral fitting
with \compmag\ model, we have fixed some of the model parameter values such as the index of velocity profile ($\eta$ = 1), 
terminal velocity at the NS surface ($\b0$ = 0.05), NS albedo A=1 and the radius of accretion column in units of the 
neutron star Schwarzschild radius as: $r_{\rm 0}$=0.25. This corresponds to $\sim$1 km for a 1.4$M_{\odot}$ mass neutron star. 
Alternatively, from the normalization constant $\left(\tfrac{R_{km}}{d_{10kpc}}\right)^{2}$ of the \compmag\ model, 
the estimated soft photons source radius $R_{km}$ varies from $\sim2.0$ to 2.8 km during the two epochs of \astrosat observations. 
We have noticed that, by increasing the value of $r_{\rm 0}$ from 0.25 to 0.5, there is no significant change in the observed 
\chisq value. The other model parameters such as the electron temperature ($kT_{e}$), the blackbody temperature ($kT_{bb}$) and
the accretion column optical depth ($\tau$) are kept free. The lesser value of $\tau$, the vertical column optical depth in 
\compmag\ model is due to electron cross-section being less than that of the Thomson cross-section $\sigma_{T}$ by a factor 
of 1000. This is due to inclusion of an energy-independent correction term which reduces $\sigma_{T}$ for photons propagated 
along the magnetic field (see \citealt{2012A&A...538A..67F} and references therein).

\section{Discussion} \label{sec:discuss}
The rapid spin changes of the slow rotation pulsar in \fourU, make it one of the interesting source to test the theories of 
spin-period evolution in long-period X-ray pulsars. In this paper, we have compared the recent changes in the pulsar spin 
period with those of previous studies. In the high mass X-ray binary system \fourU the  $\sim5500~\sec$ X-ray pulsations 
have been previously been interpreted as the NS spin period \citep{2010ApJ...709.1249F} present in the system. During 
2005--2011, the monitoring of the source with \integral has shown a long-term spin-down trend with the pulsar spin period 
from $\sim5558$ to $\sim5588~s$ \citep{2013MNRAS.432..954W}. From the \astrosat observations of the pulsar, we estimate the
present spin-period of $5619\pm3~\sec$. A recent \nustar\ observation of the source (i.e at MJD 57525.41) taken  prior to 
\astrosat observation report a spin period estimate of $5750\pm3~\sec$ \citep{2018MNRAS.479.3366T}. We note that the previous 
studies of \fourU  have confirmed the continuing long term spin-down trend of the pulsar \citep{2018MNRAS.479.3366T,
2013MNRAS.432..954W,2012MNRAS.425..595R,2009A&A...494.1073R}. However from our studies, the recent spin period estimate of the pulsar from 
\astrosat observations hints that the pulsar is showing a spin-up episode after its 20-year long spin-down process. 
We note that, the pulsar in \fourU, after reaching thus far its slowest pulse period of $\sim5750~\sec$  as seen from 
\nustar \citep{2018MNRAS.479.3366T}, has recently been transiting to a spin-up phase. Between these two epochs 
(i.e \nustar and \astrosat observations separated by $\sim 112$~days apart) we found that the pulsar in \fourU is 
transiting to a spin-up episode at a rate of $4.3\times10^{-13}$ Hz s$^{-1}$.  Before this a look at the previous 
episodes of spin-period variations in \fourU shows that,  the long term spin-down of the pulsar had continued at a rate 
$\dot{\nu}\sim-\left(1.8\pm0.1\right)\times10^{-14}$ Hz~s$^{-1} $ \citep{2018MNRAS.479.3366T}. The recent \hxmt 
observations of \fourU further confirm the significant spin-up of the slow rotating pulsar (see Fig.\ref{fig:5}). 
The resulting spin-up from the latest \nustar observation to the present \hxmt observations indicate a spin-up rate of  
$\dot{\nu} = \left(1.2\pm0.1\right)\times10^{-13} Hz~s^{-1}$ (see Figure~\ref{fig:5}). This indicates that, 
the NS undergoing a strong positive torque even at relatively low X-ray luminosity $L_{x}\leq6.5\times10^{35}$~\lum 
(see in Table~\ref{table3}). 

Such a scenario can occur in a settling quasi-spherical accretion regime \citep{2012MNRAS.420..216S}. In this regime, 
the optical companion star underfills its Roche lobe with no possibility of formation of an accretion disc around the NS. 
The radial velocity of accreting plasma is subsonic. As the net X-ray flux is low enough at the Bondi radius, the accreting 
shocked matter at the NS magnetosphere remain hot with the formation of a hot quasi-static shell. Due to temperature
gradients across the shell, large scale convective motions appear across the shell. The subsonic settling of accretion 
occurs at the magnetosphere via instabilities eventually the matter accreting onto the NS. The subsonic quasi-static shell 
mediates the removal of angular momentum from the NS  magnetosphere in settling accretion \citep{2012MNRAS.420..216S}. 

The temperature of the matter in the shell will be high  (of the order of Virial temperature, \citealt{1981MNRAS.196..209D}).
In case of \fourU, the high values of $\kte\sim 7-15$\kev imply hot matter around the NS in a quasi-static shell.
By comparing the $\kte$ values from the first and second \astrosat observations, we see a substantial Compton cooling 
of plasma. This could then facilitate the plasma to enter the NS magnetosphere from a hot envelope. Depending upon 
the sign of angular velocity difference between accreting matter and the magnetospheric boundary, the neutron star 
may spin-up or spin-down as its magnetosphere interacts with the hot envelope (or quasi-static shell). 
The settling accretion and formation of hot convective shell occurs at X-ray luminosity below a critical value of 
$\sim4\times10^{36}$\lum \citep{1981MNRAS.196..209D,2012MNRAS.420..216S} at which the angular momentum from the 
neutron star magnetosphere can be transferred away through the shell by turbulent viscosity and convective motions. 
The lower X-ray luminosity value noted in \fourU from \astrosat observations along-with higher plasma temperature, 
further confirm that, the NS underwent a settling accretion regime with possibly a short term spin-up episode.  
We observe that, the pulsar in \fourU has shown previously strong spin-down episode \citep{2018MNRAS.479.3366T} followed 
by present transition to a short-term spin-up phase during \astrosat and \hxmt observations. This suggests that, 
the pulsar is not in equilibrium. 

Rapid spin-up episodes on top of the long-term spin-down trend have also been observed in other long period 
pulsars such as \gx\ \citep{2001ApJ...554..383P,2010A&A...515A..10D,2021MNRAS.506.2712D}. These changes reflect 
that the pulsar is entering into a strong coupling regime \citep{2012MNRAS.420..216S} temporarily. During which, 
the accreting matter at the Alfv\'en radius ($R_{A}$) couples with the magnetosphere acquiring angular velocity 
of the neutron star. The accreting matter as it falls onto the NS, it returns the angular momentum to the NS 
acquired at $R_{A}$ via the strong magnetic field. This leads to a significant spin-up of the pulsar. 
During the recent spin-down state of the pulsar, \citet{2018MNRAS.479.3366T} have obtained a lower limit on the 
NS dipole magnetic moment of the pulsar to be $\mu>10^{31}$~G~cm$^{-3}$, where they have assumed the pulsar 
is in spin-equilibrium with the X-ray luminosity much smaller than the equilibrium luminosity, neglecting 
the spin-up torque at the spin-down stage. The dipole magnetic field of a wind-fed  NS as per the spherical 
settling accretion theory \citep{2012MNRAS.421L.127P}, can be estimated as:
\begin{equation} \label{eq:2}
    B_{12} \sim 8.1\, \dot M_{16}^{1\over 3} V_{300}^{-{11\over 3}}
    \left(\frac{P_{1000}}{P_{orb,300}}\right)^{11 \over 12},
\end{equation}
where the mass accretion rate $\dot M_{16}$ is in units of $10^{16}~gs^{-1}$, the stellar wind velocity $V_{300}$ is in units of 300 km s$^{-1}$, the spin-period, $P_{1000}$ is in units of 1000~sec and the orbital period is in units of 300~d as $P_{orb,300}$. 
 
From the \astrosat observed source luminosity of \fourU in 1-60\kev, \ie $L_{x}\sim6.5\times10^{35}$\lum, 
we obtain the mass accretion rate of $ \dot M_{16}\sim0.35$.  The stellar wind velocity in \fourU is very 
slow with $V_{300}\sim1.17$ \citep{2006A&A...449..687R, 2012MNRAS.425..595R}. Using the above values and the 
present spin-period estimate of the pulsar from \hxmt observations as, $P_{1000} = 5.291$, we obtain an 
estimate of the pulsar magnetic field in \fourU using equation~\ref{eq:2} as: $B = 3.6\times10^{14}~G$ 
for an orbital period of $P_{orb,300}=0.031$ \citep{2006A&A...449..687R} and alternatively 
$B = 1.85\times10^{14}~G$ if we consider $P_{orb,300}=0.064$. The above two possible binary orbital modulations 
in case of \fourU were considered due to \citet{2006A&A...449..687R} found $P_{orb}=9.5d$ 
and later a $19d$ modulation was suggested by \citet{2007ApJ...655..458C}. Nevertheless, it is interesting 
to note that, with an orbital period of 9.5d, \fourU has the shortest orbital period among the known HMXBs 
\citep{2022MNRAS.511.4123H}. Considering the NS surface magnetic field as 
$B = 2\mu/R^{3} \geq 1.85\times10^{14}~G$, where $R$ is the NS radius and $\mu$ is NS dipole 
magnetic moment. we obtain $\mu>9.25\times10^{31}~G~cm^{3}$. These estimates agree with the results obtained 
in \citet{2018MNRAS.479.3366T}, where the spin-up torque has been ignored while estimating the NS dipole magnetic 
moment in a quasi-spherical settling accretion scenario. 

As per the quasi-spherical settling accretion theory, the equilibrium period of an X-ray pulsar is determined as
\citep{2018ASSL..454..331S}: 
\begin{equation}
      P_{eq} = 1000\,[s]\,\mu_{30}^{12/11}\left(\frac{P_{b}}{10d}\right)\dot M_{16}^{-{4/11}}V_{8}^{4},
\end{equation}
where the NS dipole magnetic moment $\mu_{30}$ is in units of $10^{30}$~\gcmq, and the stellar wind velocity $v_{w}$ is in units of $10^{8}$ cm s$^{-1}$ or \ie $V_{8}$ in units of 1000 km s$^{-1}$. For a comparison, in case of \fourU, using $v_{w}=350$ km s$^{-1}$ \ie $V_{8}=0.35$ \citep{2006A&A...449..687R} and $P_{orb}=9.5$~d, 
$\mu_{30}=9.25\times10^{31}$~\gcmq (as obtained above), we find the pulsar equilibrium period is of 
$P_{eq}=2823$~s. Whereas in a standard accretion disk accretion case, the equilibrium period 
\citep{2018ASSL..454..331S} would be determined from :
\begin{equation}
P_{eq,d} \approx 10[s]\mu_{30}^{6/7}{\dot M_{16}}^{-3/7}.
\end{equation}
In case of \fourU, it is estimated to be $P_{eq,d}=760~s$.  In either scenario, it clearly demonstrates the 
fact that, \fourU is a non-equilibrium pulsar. The lack of any CRSF feature in the hard X-ray spectra of 
\fourU from the \astrosat observations, along with the indication of high-magnetic field of the pulsar, 
$B>10^{14}~G$ as shown above, (see also \citealt{2018MNRAS.479.3366T,2013MNRAS.432..954W,2012MNRAS.425..595R}) 
hints at a magnetar field strength of the long-period pulsar in \fourU.

\section{Summary and Conclusion} \label{sec:summary}
We have studied the recent spin-period evolution and spectral changes in the long-period pulsar in \fourU 
using \astrosat and \hxmt observations. These studies reveal that, the pulsar is showing a rare spin-up 
episode, with a short-term spin-up rate of $\dot\nu=4.3\times10^{-13}$~Hz~$s^{-1}$ from \nustar to \astrosat 
observations. The net spin-up rate from \nustar to \hxmt observation gradually decreases to 
$\dot\nu=1.2\times10^{-13}$~Hz~$s^{-1}$. The phase-averaged spectral studies with Comptonization models such 
as \comptt and \compmag reveal a hotter soft-photon source region varying in size of $2-2.8$~km near the pulsar. 
The net X-ray luminosity of the pulsar in 1-60\kev is $\leq6.5\times10^{35}$~\lum as seen from \astrosat. 
In a quasi-spherical settling accretion scenario, this results in a formation of hot convective and quasi-static
shell around the NS. The convective motions in the shell regulates the angular momentum transfer from the 
NS magnetosphere that may result in a spin-up of the pulsar. A comparison of present spin-period value with 
the equilibrium spin period of the pulsar obtained in different accretion regimes shows that pulsar is not 
in equilibrium. The non-detection of any CRSF feature in 1-60~\kev along-with the possibility of strong magnetic 
field NS in \fourU as per quasi-spherical settling accretion imply the NS of magnetar field strength in \fourU. 
Given the low X-ray luminosity of the pulsar and slow and marginal pulsations, the present spin-up episode  
can be better explored with future observations with the hard X-ray telescopes like \nustar, \astrosat and \hxmt. 

\section*{Acknowledgments}
We are grateful to the referee for the detailed comments. This work is supported by the the NSFC (No. 12133007) and National Key Research and Development Program of China (Grants No. 2021YFA0718503). This work has made use of data from the Insight-HXMT mission, 
a project funded by China National Space Administration (CNSA) and the Chinese Academy of Sciences (CAS). This work has performed utilizing the calibration data-bases and auxillary analysis tools provided by AstroSat-SXT team and also used the data from the Soft X-ray Telescope (SXT) developed at TIFR, Mumbai.




\bibliography{4u2206_ms}{}
\bibliographystyle{aasjournal}

\label{lastpage}

\end{document}